\documentclass[
pra,twocolumn,showpacs,preprintnumbers,amsmath,amssymb]{revtex4}
\pdfoutput=1
\usepackage{graphicx}
\usepackage{amssymb}
\usepackage{amsmath}
\usepackage{dsfont}
\usepackage{bm}
\usepackage{mathrsfs}
\usepackage{times}

\usepackage{feynmf}
\usepackage{amsfonts}

\allowdisplaybreaks[1]

\begin{document}
\title{Absorption in dipole-lattice models of dielectrics}

\author{R.\ J.\ Churchill}

\email{rc313@exeter.ac.uk}

\author{T.\ G.\ Philbin}

\email{t.g.philbin@exeter.ac.uk}

\affiliation{Physics and Astronomy Department, University of Exeter,
Stocker Road, Exeter EX4 4QL, United Kingdom}

\begin{abstract}
We develop a classical microscopic model of a dielectric. The model features nonlinear interaction terms between polarizable dipoles and lattice vibrations. The lattice vibrations are found to act as a pseudo-reservoir, giving broadband absorption of electromagnetic radiation without the addition of damping terms in the dynamics. The effective permittivity is calculated using a perturbative iteration method and is found to have the form associated with real dielectrics. Spatial dispersion is naturally included in the model and we also calculate the wavevector dependence of the permittivity.
\end{abstract}

\maketitle

\section{Introduction}

Macroscopic electromagnetism remains a central component in describing a large variety of interactions between light and matter. The electromagnetic response of macroscopic materials is routinely captured by an electric permittivity, and the permittivity itself can often be fitted to a simple function of frequency that belies the complexity of the underlying microscopic physics. Typically the permittivity $\varepsilon(\omega)$ of a dielectric is expected to take the form $1-\sum_n\alpha_n/(\omega^2-\omega_n^2+i\gamma_n\omega)$, corresponding to a series of broadened resonances with attendant absorption~\cite{jac}. A simple argument~\cite{jac}, based on a polarizable particle with damping together with the assumption of a rarified material, leads to this formula for the permittivity. It is  however difficult to justify this result with a more realistic model~\cite{Hopfield}. In particular, when one requires the absorption of light to emerge from the model without being put in by hand as a damping term, one faces some difficult calculations~\cite{Hopfield}. A classic paper by Hopfield in 1958~\cite{Hopfield} elucidated the connection between the interaction of light with a polarizable material on the one hand, and an effective description of such interactions by a permittivity on the other hand. Hopfield proposed that a rather simple model of the light-matter interaction would lead to the permittivity given above, without the need to impose damping by hand or assume a rarified material. Although Hopfield's model is simple to state and express as a Lagrangian, the derivation of the resulting effective permittivity requires considerable effort and does not appear to have been carried out before. The aim of this paper is to verify Hopfield's conjecture and demonstrate that the formula for $\varepsilon(\omega)$ above can be derived from a simple, intuitive, classical model of a dielectric. 

The initial model considered by Hopfield in~\cite{Hopfield} is that of a continuous polarization field linearly coupled to light. This does not give realistic results however because light is only absorbed at the resonant frequency of the polarization field, giving a permittivity that has a delta function in frequency as its imaginary part~\cite{Hopfield}. The problem with this model is identified by Hopfield as an insufficient density of final states for energy transitions from light into the matter degrees of freedom. Hopfield states that the real reason for absorption of light by materials is nonlinear interaction between the electric dipole moments induced by light and lattice vibrations, so his polarization field would have to be nonlinearly coupled to lattice vibrations in order to see realistic absorption behaviour. No calculations however are given in~\cite{Hopfield} for such a model with nonlinear matter interactions. Subsequently Hopfield's suggestion has been followed up in a quantum setting where excitons (the quantized polarization field) are nonlinearly coupled to lattice phonons, the latter providing an effective damping when light interacts with the former~\cite{Tait,Hiz,Mav,Egl,Cam}. A full analysis in the quantum setting is very challenging whereas an appealing aspect of Hopfield's proposal is that it can be carried out classically, much as the textbook motivations~\cite{jac} for the macroscopic Maxwell equations are classical. Here we will clarify that the broad absorption of the form used in the standard permittivity expressions can be derived from elementary classical physics.

 In Hopfield's scenario the lattice vibrations act as a reservoir into which electromagnetic energy can dissipate. This subsequently led to another approach in which a phenomenological reservoir consisting of a field of harmonic oscillators at every frequency is \emph{linearly} coupled to the polarization field~\cite{Huttner,Suttorp}. This phenomenological reservoir (an uncountable continuum of harmonic oscillators) is meant to mimic the dissipative channel into lattice vibrations through nonlinear interactions proposed by Hopfield. The continuum reservoir, introduced by Huttner and Barnett~\cite{Huttner}, allows dissipation and a realistic permittivity to be derived from a linear model but a detailed connection to microscopic physics is less clear. The continuum reservoir approach has proven to be a powerful tool in  incorporating dissipation in a Lagrangian with linear coupling. It has been used to give a Lagrangian formulation of the macroscopic Maxwell equations for an arbitrary permittivity, without use of a polarization field~\cite{bha06,khe06,sut07,amo08,amo09,khe10,philbin2010canonical}, and similarly to give a Lagrangian formulation of damped harmonic oscillators generally~\cite{philbin2012quantum}. This gives a classical and quantum description of light in all absorbing and dispersive macroscopic media, where the required permittivity (and magnetic permeability) are incorporated through coupling functions in the Lagrangian. Here we return to the considerations~\cite{Hopfield}  that led to the reservoir approach, to further clarify the microscopic processes that allow the macroscopic behaviour to be so accurately captured by a phenomenological reservoir. We seek to verify that linear coupling of light to electric dipoles which are in turn nonlinearly coupled to lattice vibrations leads to a permittivity well described by the simple textbook formula given above, a permittivity that in particular exhibits the broadband absorption characteristic of real materials. Our analysis will also capture the nonlocal response (spatial dispersion) of the material medium, in addition to its temporal dispersion, giving insight into the wavevector dependence of the permittivity of dielectrics.
 
 Because Hopfield's proposal views the material as a lattice of dipoles, the model considered here is applicable to non-metallic solid dielectrics with a regular lattice. The model is thus not appropriate for amorphous materials, liquids or gasses, though in practice the permittivity functions of all these states of matter often show similar features.
 
Although our motivation is to elucidate the microscopic physics of natural materials, it is of course possible to construct experimentally a lattice of dipoles on macroscopic scales. Metamaterials based on arrays of dielectric nanoparticles provide an example~\cite{zha09,sou11}. Such artificial materials are being actively investigated as low-loss alternatives to structures with metallic components~\cite{sch07,wan14,jia16}. The internal structure of the  dielectric nanoparticles, however, is an important aspect of such materials and nothing corresponding to this internal structure is included in our model. (An important aspect of such dielectric nanoparticles is that the size of the particles can be used to control a magnetic as well as an electric response~\cite{sch07}.) The coupling between the nanoparticles in the array is important but usually this coupling is investigated for its effect in altering the resonance structure of the metamaterial~\cite{wan14,jia16}. In contrast, the important effect of the dipole coupling in our model is to induce the lattice vibrations that are essential for obtaining broadband absorption in our calculations. On the other hand, in the microwave regime a lattice of dipoles that can vibrate is feasible in principle and may be useful to explore the absorption and non-local response captured by our model.

The outline of the paper is as follows. In  Sec.~\ref{sec:linear} we consider a simple 1D microscopic linear model and derive an expression for the permittivity. We discuss the failings of such a model and the need for a nonlinear interaction term as a pseudo-reservoir.
This is added to the model in Sec.~\ref{sec:nonlinear} and the equations of motion are solved in Sec.~\ref{sec:perturbative} using a perturbative iteration procedure.
In Sec.~\ref{sec:diagrams} we introduce a graphical representation of the perturbative solution, which is used in  Sec.~\ref{sec:effectiveperm} to find the effective permittivity of the medium. Numerical calculations for the frequency and wavevector dependence of the effective permittivity are given in Sec.~\ref{sec:numerical} and \ref{sec:spatial}.

\section{Linear model of a dielectric}   \label{sec:linear}
We initially consider a simple 1D linear model that will prove inadequate to capture the absorption of light by materials. This will demonstrate why nonlinear interactions need to be incorporated into the model.

Consider an infinite chain of polarizable dipoles positioned at $x_n=na$, where $a$ is the lattice spacing. 
This model can also be considered as the limit of the large but finite chain, described by the same calculations with some small approximations.
Each dipole moment is modelled as a harmonic oscillator $p_n(t)$, with a Lagrangian:
\begin{align}
L_{p}=\frac{1}{2}\sum_{n=-\infty}^\infty
\bigg[&
\dot{p}^2_n 
-\omega_0^2 p^2_n 
\nonumber\\
&
+\frac{1}{2}
\sum_{m\ge1}^\infty\tau_m
\left(
p_np_{n-m}
+
p_np_{n+m}
\right)
\bigg],
\end{align}
\noindent where $\omega_0$ is the resonant frequency of the harmonic oscillators and $\tau_{m}$ describes the dipole-dipole interaction for a symmetric and translationally invariant system.
This model of a dielectric can be viewed as a discrete version of the Hopfield model of a continuous polarization field~\cite{Hopfield}. 
The second part of our model is a 1D scalar field $\phi(x,t)$, representing a projection of the vector potential $\bf{A}$, with the Lagrangian:
\begin{equation}
L_\phi=\frac{1}{2}\int_{-\infty}^\infty dx \left[\frac{1}{c^2}\dot{\phi}^2(x,t) -(\partial_x\phi(x,t))^2\right],
\end{equation}
\noindent where $c$ is the speed of light.
The dipole moment is coupled to the time derivative of the scalar field, representing the electric field $\bf{E}=-\bf{\dot{A}}$.
An additional feature is a spatially dependent coupling term between the dipole moment $p_n$ and the field near the lattice site $x_n$. The function $\alpha(x-x_n)$ is used to account for the finite size of the dipoles, which are the ``atoms" in our model. The interaction Lagrangian is given by:
\begin{equation}
L_{\phi p}=
-d_0
\sum_{n=-\infty}^\infty
p_n\int_{-\infty}^\infty dx\, \alpha(x-na)\dot{\phi}(x),
\end{equation}
\noindent where $d_0$ is the coupling strength and $\alpha(x)$ is taken to be a normalized Gaussian function, with the convenient feature that in the limit $\sigma\to0$ it becomes a Dirac delta function:
\begin{equation}
\alpha(x)=\frac{1}{\sigma\sqrt{2\pi}}e^{-\frac{x^2}{2\sigma^2}}.
\end{equation}

At this point we make a spatial Fourier transformation. For the field, this is given by:
\begin{equation}
\phi(x)=\frac{1}{\sqrt{2\pi}}\int_{-\infty}^\infty dk\,\phi(k)e^{ikx},
\end{equation}
\noindent where $k$ is a continuum over all values. For the infinite medium, the following expressions are used:
\begin{gather}
p_n=
\sqrt{\frac{a}{2\pi}}
\int_{-\pi/a}^{\pi/a}dq\,
p(q)e^{iqan},   \label{pnFourierrep}
\\
\sum_{n=-\infty}^\infty e^{iqan}
=\frac{2\pi}{a}
\sum_{j=-\infty}^\infty
\delta\left(q+j\frac{2\pi}{a}\right),
\label{eq:reciprocal_lattice_scattering}
\end{gather}
\noindent where $q$ is now a continuous variable over the first Brillouin zone $-\tfrac{\pi}{a}<q<\tfrac{\pi}{a}$.
The delta functions in (\ref{eq:reciprocal_lattice_scattering}) contain terms where $q$ is displaced by an integer number of the reciprocal lattice vector $2\pi/a$.
As we are primarily interested in initial fields with $\lambda\gg a$, we only consider the $j=0$ term at this point.
Due to the real nature of the initial variables, the transformed variables obey $\phi(-k)=\phi^*(k)$. The Lagrangian for this system now takes the form:
\begin{gather}
L_{\phi}=\frac{1}{2}\int_{-\infty}^{\infty}dk\left[\frac{1}{c^2}\dot{\phi}(k)\dot{\phi}(-k)-k^2\phi(k)\phi(-k)\right],
\\
L_{p}=
\frac{1}{2}
\int_{-\frac{\pi}{a}}^{\frac{\pi}{a}} dq\,
\left[\dot{p}(q)\dot{p}(-q)- \omega_0^2(q) p(q)p(-q)\right],
\\
L_{\phi p}=
-d_0
\sqrt{\frac{2\pi}{a}}
\int_{-\infty}^{\infty}dk
\int_{-\frac{\pi}{a}}^{\frac{\pi}{a}} dq\,
p(q)
\alpha(-k)
\dot{\phi}(k)
\delta(k+q)
,
\end{gather}
\noindent where $\omega_0(q)$ is the dipole dispersion equation in the absence of $L_{\phi p}$ and calculated from $\tau_m$:
\begin{align}
\omega^2_0(q)
=
&
\omega_0^2
-
\frac{1}{2}
\sum_{m\ge1}^\infty
\tau_m
\left(
e^{-iqam}
+
e^{iqam}
\right)
\nonumber\\
=&
\omega_0^2
-
\sum_{m\ge1}^\infty
\tau_m
\cos(qam),
\label{eq:resonant_frequency}
\end{align}
\noindent where we will only consider nearest neighbour coupling, with $\tau_{m\ge2}=0$.
The equations of motion of the Lagrangian are:
\begin{equation}
\ddot{\phi}(k,t)+(ck)^2\phi(k,t)=d_0c^2\sqrt{\frac{2\pi}{a}}\alpha(k)\dot{p}(k,t),
\end{equation}
\begin{equation}
\ddot{p}(k,t)+\omega_0^2(k)p(k,t)=
-d_0
\sqrt{\frac{2\pi}{a}}
\alpha(-k)\dot{\phi}(k,t).
\end{equation}
\noindent Using the Fourier transformation:
\begin{equation}
\phi(t)=\frac{1}{\sqrt{2\pi}}\int_{-\infty}^\infty d\omega\, \phi(\omega)e^{-i\omega t},
\end{equation}
\noindent with the property $\phi(-k,-\omega)=\phi^*(k,\omega)$, the equations of motion become:
\begin{equation}
\label{eq:linear_phi_equation}
\left[k^2-\left(\frac{\omega}{c}\right)^2\right]
\phi(k,\omega)=
i\omega \frac{d_0}{\sqrt{a}}
\sqrt{2\pi}\alpha(k)
p(k,\omega),
\end{equation}
\begin{equation}
\label{eq:linear_p_equation}
\left[\omega^2_0(k)-\omega^2\right]p(k,\omega)=
-i\omega \frac{d_0}{\sqrt{a}}
\sqrt{2\pi}\alpha(-k)
\phi(k,\omega).
\end{equation}

Solving \eqref{eq:linear_p_equation} gives:
\begin{equation}
p(k,\omega)=p_h(k,\omega)
-i\omega \frac{d_0}{\sqrt{a}}
\sqrt{2\pi}\alpha(-k)
G_p(k,\omega)\phi(k,\omega),  \label{psollin}
\end{equation}
\noindent where $p_h$ is the homogenous solution of $p$ satisfying the equation of motion in the absence of coupling:
\begin{equation}
\left[\omega^2_0(k)-\omega^2\right]p_h(k,\omega)=0,
\label{eq:p_h}
\end{equation}
and the retarded Green function $G_p$ takes the form
\begin{align}
G_p(k,\omega)=&
\frac{1}{\omega_0^2(k)-(\omega+i0^+)^2}
\nonumber\\
=&
P\frac{1}{\omega_0^2(k)-\omega^2}
+
\nonumber\\&
\frac{i\pi}{2\omega_0(k)}
\left[
\delta(\omega-\omega_0(k))
-\delta(\omega+\omega_0(k))
\right]
,
\label{eq:G_p_expansion}
\end{align}
\noindent where the pole at $\omega^2=\omega^2_0(k)$ has been moved into the lower-half complex plane by introducing the infinitesimal positive value $0^+$. This ensures that the solution is dependent on the field at previous times and satisfies the Kramers-Kronig relations.  Substituting (\ref{psollin}) into \eqref{eq:linear_phi_equation} gives
\begin{equation}
\left[
k^2-
\frac{\omega^2}{c^2}
\varepsilon(k,\omega)
\right]
\phi(k,\omega)
=i\omega \frac{d_0}{\sqrt{a}}
\sqrt{2\pi}\alpha(k)
p_h(k,\omega),
\label{eq:wave_equation}
\end{equation}
\noindent where the relative permittivity $\varepsilon$ is given by:
\begin{align}
\varepsilon(k,\omega)
=&1+\frac{d_0^2c^2}{a}\left(2\pi|\alpha(k)|^2\right)
G_p(k,\omega)
\nonumber\\
=&
1+
\frac{
\left(
d_0^2c^2/a
\right)
\left(2\pi|\alpha(k)|^2\right)
}{
\omega^2_0(k)
-
\left(
\omega 
+i0^+
\right)^2
}
.
\label{eq:permittivity}
\end{align}

Equation \eqref{eq:G_p_expansion} shows that for a given $k$, the imaginary part of the permittivity is given by a Dirac delta function at the corresponding resonant frequency $\omega_0(k)$, when the mode $(k,\omega)$ of the field $\phi$ lies on the $p$ dispersion relation.
As $\omega_0(k)$ is periodic in $k$, $\alpha(k)$ ensures $\varepsilon\to1$ as $k\to\infty$ for any non-Dirac delta function spatial coupling. As noted above, we only included the $j=0$ term in (\ref{eq:reciprocal_lattice_scattering}) so the result (\ref{eq:permittivity}) is not the exact solution. Nevertheless the exact expression for the permittivity also has an imaginary part that is a delta function. 

In reality, the complex permittivity near a resonant frequency has a finite imaginary component over a range of frequencies.
This behaviour is usually modelled by treating each dipole as a damped harmonic oscillator (DHO), modifying $G_p$ to the form:
\begin{equation}
G_p(k,\omega)=\frac{1}{\omega_0^2(k)-\omega^2-i\gamma\omega},  \label{GpDHO}
\end{equation}
\noindent where $\gamma$ is a damping term. However, recovering the DHO equations of motion from a Lagrangian 
presents challenges. The oscillator can be coupled to either a discrete~\cite{mag59,fey63,cal83,Tatarskii} or a continuous~\cite{philbin2012quantum} phenomenological reservoir. The oscillator-reservoir coupling must however take a very specific form in both cases if damping of the form seen in (\ref{GpDHO}) is to be recovered. In practice this damping term is usually added by hand to calculations. As this kind of damping leads to the standard permittivity expression stated at the beginning of this paper, we wish our model to produce it from physically motivated interactions, without extreme fine tuning of coupling terms in the Lagrangian.

Hopfield faced the same problem of infinite absorption at a single resonant frequency in a similar dielectric model \cite{Hopfield}, consisting of the electromagnetic field linearly coupled to a continuous harmonic-oscillator field (a polarization field, or exciton field in quantum language).
He identified the problem as due to the linear coupling, which allows only coupling between single modes due to wavevector conservation. In terms of second-order perturbation theory,  the $photon\to exciton$ process does not have a density of final states, and so no real transitions occur.
He then suggests that three-body (and higher order) interactions are responsible for absorption, focusing on the $exciton\to exciton^\prime+phonon$ process.
Unlike the linear case, there is an additional degree of freedom in interactions, with a continuum of coupled modes with the same total wavevector. This gives a continuum of final states with real transitions as energy absorbed from the electromagnetic field is stored in the indirectly coupled phonons.
The coupling of the excitons to a pseudo-reservoir with the same wavevector but a range of energies was part of the motivation behind the Huttner-Barnett model~\cite{Huttner}.
Hopfield states~\cite{Hopfield} that such nonlinear interactions give rise to an effective damping term such as appears in (\ref{GpDHO}), however no calculations are given. We now consider such interactions in a classical model.

\section{Nonlinear model of a dielectric}   \label{sec:nonlinear}
\begin{figure*}[htb]
\centering
{\includegraphics[width=170mm]{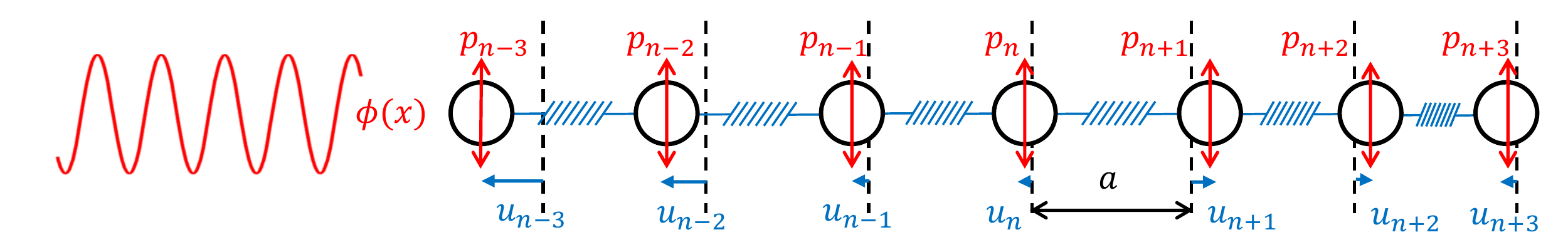}}
\caption{
Schematic of the nonlinear model of a dielectric. The dipole moments $p_n$ are linearly coupled to the scalar field $\phi$ and nonlinearly coupled to the atomic displacement $u_n$ from the lattice site $x_n=na$. In addition, the atomic displacements are coupled to each other, with nearest-neighbour coupling displayed here.
}
\label{fig:model}
\end{figure*}

We modify the model of the previous section by expanding the behaviour of the medium into two degrees of freedom: the dipole moments of the ``atoms" $p_n$ and the physical displacement of the atoms from the equilibrium positions $x_n$ given by $u_n$ (see Fig.~\ref{fig:model}).

The atom displacement is not directly coupled to the field, instead being indirectly coupled through a nonlinear interaction with the dipole moments. This nonlinear coupling arises naturally from the $r^{-3}$ dipole-dipole interaction:
\begin{align}
L_{pp}=&
V_0
\sum_{n=-\infty}^\infty
\sum_{m=-\infty,\ne0}^\infty
\frac{p_np_{n-m}}
{\left|
(x_n+u_n)
-
(x_{n-m}+u_{n-m})
\right|^3
}
\nonumber\\
=&
V_0
\sum_{n=-\infty}^\infty
\sum_{m=-\infty,\ne0}^\infty
\frac{1}{\left|ma
\right|^3}
\frac{p_np_{n-m}}
{
\left|
1+\tfrac{u_n-u_{n-m}}{ma}
\right|^3
}.
\label{eq:expansion}
\end{align}
Performing a small $u$ expansion gives a series of Lagrangian terms with increasing powers of $u$. The first term is linear and is included in $L_p$ with $\tau_{m}$. The first nonlinear term, considering only nearest neighbour $|m|=1$ interactions, is
\begin{align}
L_{ppu}=
V_0
\sum_{n=-\infty}^\infty
\frac{-3}{a^3}
\bigg[&
p_np_{n-1}
\left(
\tfrac{u_n-u_{n-1}}{a}
\right)
\nonumber\\&
-
p_np_{n+1}
\left(
\tfrac{u_n-u_{n+1}}{a}
\right)
\bigg]
,    \label{Lppuxspace}
\end{align}
In addition, the atomic displacements are linearly coupled to each other by a Lagrangian term $L_u$:
\begin{align}
L_{u}=&\frac{1}{2}m_0\sum_{n=-\infty}^{\infty}
\bigg[
\dot{u}^2_n- 
\kappa_0u_n^2
\nonumber\\
&-
\frac{1}{2}
\sum_{m\ge1}^\infty\kappa_m
\left(
u_nu_{n-m}
+
u_nu_{n+m}
\right)
\bigg],
\end{align}
\noindent where $m_0$ is the mass of the atom and $\kappa_{m}$ are the interaction terms. After a spatial Fourier transform:
\begin{equation}
L_{u}=
\frac{1}{2}m_0
\int_{-\frac{\pi}{a}}^{\frac{\pi}{a}} dq\,
\left[\dot{u}(q)\dot{u}(-q)- \Omega^2(q) u(q)u(-q)\right],
\end{equation}
\noindent where the new resonant frequency is calculated in the same manner as $\omega_0(q)$ in \eqref{eq:resonant_frequency}:
\begin{equation}
\Omega^2(q)
=
\kappa_0
-
\sum_{m\ge1}^\infty
\kappa_m
\cos(qam),
\label{eq:u_resonant_frequency}
\end{equation}
\noindent and we again only consider nearest-neighbour interactions with $\kappa_{m>2}=0$.
The nonlinear Lagrangian term (\ref{Lppuxspace}) takes the following form in Fourier space:
\begin{widetext}
\begin{gather}
L_{ppu}=
\frac{
\gamma
}{
\sqrt{2\pi a}
}
\int_{-\frac{\pi}{a}}^{\frac{\pi}{a}}
\int_{-\frac{\pi}{a}}^{\frac{\pi}{a}}
\int_{-\frac{\pi}{a}}^{\frac{\pi}{a}}dq_1dq_2dq_3
\left[
f_0(q_1,q_2,q_3)
p(q_1)p(q_2)
u(q_3)
\delta(q_1+q_2+q_3)
\right]
,    
\\
f_0(q_1,q_2,q_3)=i[\sin(q_2a)-\sin((q_2+q_3)a)],
\end{gather}
\noindent where $\gamma=6V_0/a^3$ and we have omitted reciprocal lattice vector terms present due to  \eqref{eq:reciprocal_lattice_scattering} for notational brevity.
After evaluating the Dirac delta functions, the equations of motion become:
\begin{align}
\phi(k,\omega)=&\phi_h(k,\omega)+i\omega \frac{d_0}{\sqrt{a}}\left(\sqrt{2\pi}\alpha(k)\right)G_\phi(k,\omega)p(k,\omega),
\label{eq:nonlinear_phi_equation}
\\
p(k,\omega)=&p_h(k,\omega)
-i\omega \frac{d_0}{\sqrt{a}}\left(\sqrt{2\pi}\alpha(-k)\right)G_p(k,\omega)\phi(k,\omega)
\nonumber\\
&+\gamma 
\left(
\frac{c}{2\pi a^{5/2}}
\right)
G_p(k,\omega)
\int_{-\pi}^{\pi} d(q_1a)
\int_{-\infty}^\infty d\left(\tfrac{\omega_{q_1}a}{c}\right)
\bigg[
f_1(k,q_1)
u(q_1,\omega_{q_1})
p(k-q_1,\omega-\omega_{q_1})
\bigg],
\label{eq:nonlinear_p_equation}
\\
u(q,\omega)=&
u_h(q,\omega)
+
\gamma 
\left(
\frac{c}{2\pi a^{5/2}}
\right)
G_u(q,\omega)
\int_{-\pi}^{\pi} d(q_1a)
\int_{-\infty}^\infty d\left(\tfrac{\omega_{q_1}a}{c}\right)
\bigg[
f_2(q,q_1)
p(q_1,\omega_{q_1})
p(q-q_1,\omega-\omega_{q_1})
\bigg],
\label{eq:nonlinear_u_equation}
\end{align}
\end{widetext}
\noindent where the integration variables have been rescaled to dimensionless values and
\begin{gather}
f_1(k,q_1)=
-2i[\sin(ka)-\sin((k-q_1)a)],
\label{eq:f1}
\\
f_2(q,q_1)=
i[\sin((q-q_1)a)-\sin(qa)].
\end{gather}
We have also introduced the retarded Green functions $G_\phi$ and $G_u$, given by
\begin{gather}
G_\phi(k,\omega)=\frac{c^2}{(ck)^2-(\omega+i0^+)^2},
\\
G_u(q,\omega_q)=\frac{1/m_0}{\Omega^2(q)-(\omega+i0^+)^2},
\end{gather}
\noindent and the homogenous solutions of the fields $\phi_h$ and $u_h$ satisfy the equations
\begin{gather}
\left[\omega^2-(ck)^2\right]\phi_h(k,\omega)=0,
\\
\left[\omega_q^2-\Omega^2(q)\right]u_h(q,\omega_q)=0.
\label{eq:homogenous_u_equation}
\end{gather}
Equations \eqref{eq:p_h} and \eqref{eq:G_p_expansion} for $G_p$ and $p_h$ remain unchanged. 
Here we note that the nonlinear terms have introduced a pseudo-reservoir to the $p$ equation of motion (\ref{eq:nonlinear_p_equation}), where a single initial mode is coupled to a continuum of modes with the same total $k$ and $\omega$. 

\section{Perturbative Solution}   \label{sec:perturbative}

We now present a method of calculating an effective permittivity for this nonlinear system by deriving a wave equation for $\phi$ similar to \eqref{eq:wave_equation}.
This is done using an iteration procedure, treating terms with small nonlinear coupling coefficient $\gamma$ as a perturbation of the linear model.
We first consider \eqref{eq:nonlinear_u_equation}, as $u$ is not directly coupled to $\phi$.
In our model, the expression for $u(q,\omega)$ can be immediately substituted into \eqref{eq:nonlinear_p_equation} as it is expressed solely in terms of the homogenous solution $u_h$ and $p$.
For a system with different nonlinear coupling, the right-hand side (RHS) of \eqref{eq:nonlinear_u_equation} may contain additional terms of $u(q^\prime,\omega^\prime)$. For example, further expansion of \eqref{eq:expansion} gives a $ppuu$ term in the Lagrangian and a $ppu$ term in \eqref{eq:nonlinear_u_equation}.
In this case all RHS $u$ terms in \eqref{eq:nonlinear_u_equation} are repeatedly iterated. 
After $n$ iterations, all RHS terms up to the $n^{th}$ power of $\gamma$ contain only $u_h$ and $p$, while terms still involving $u$ are of the order $\gamma^{n+1}$ or higher.
Removing the remaining $u$ terms leaves an expression for $u(q,\omega)$, accurate up order $\gamma^n$.

The same situation is found upon substituting this expression into \eqref{eq:nonlinear_p_equation}, with $p(k,\omega)$ in terms of homogenous solutions $(u_h, p_h)$, the field $\phi$ and additional RHS terms of $p(k^\prime,\omega^\prime)$.
These terms are repeatedly iterated using the new equation of motion for $p$, to give an expression solely in terms of $u_h$, $p_h$ and $\phi$ accurate up to order $n$ in $\gamma$.

A nonlinear wave equation for $\phi$ is found upon substitution of $p(k,\omega)$ into \eqref{eq:nonlinear_phi_equation}. The RHS terms can be split into three groups: those containing only homogenous solutions ($u_h$, $p_h$ and $\phi_h$), those linear in $\phi$ and those nonlinear in $\phi$.
The nonlinear $\phi$ terms can be used to find an effective nonlinear permittivity; these terms can also be used to analyse the re-emission of frequency-converted $\phi$ waves from an absorbed incident beam.
These nonlinear terms start with higher-order powers of $\gamma$ compared to the linear terms and are dependent on higher powers of $\phi$. As a result, we can consider these terms to be negligible for ``weak'' fields.
This leaves the terms linear in $\phi$.
After substituting $p(k,\omega)$ into \eqref{eq:nonlinear_phi_equation}, the linear $\phi$ terms up to $\gamma^n$ are:
\begin{align}
\left[k^2-\left(\frac{\omega}{c}\right)^2\right]
\phi(k,\omega)=
&
\omega^2
\frac{d_0^2}{a}\left(2\pi|\alpha(k)|^2\right)
G_p(k,\omega)
\phi(k,\omega)
\nonumber\\
&+Z\left[
\phi
\right]
+
O\left[\gamma^{n+1}\right]
\label{eq:linear_phi_terms}
\end{align}
\noindent where $Z[\phi]$ is a linear functional of  $\phi$, containing an integration $\int dk^\prime \int d\omega^\prime$ over  $\phi(k^\prime,\omega^\prime)$ terms. 

To find an effective permittivity of the medium from \eqref{eq:linear_phi_terms}, we perform a slightly different iteration procedure to the previous two equations of motion. 
The linear $\phi$ terms in  $Z[\phi]$ can be further split into two groups: those in the same mode $(k,\omega)$ as the other $\phi$ terms in (\ref{eq:linear_phi_terms}) and those in a different mode $(k'\neq k,\omega'\neq\omega)$. The terms in the mode $(k,\omega)$ however are a set of measure zero in an  integration $\int dk^\prime \int d\omega^\prime$ over all modes. This is a result of the continuous nature of $q$ in (\ref{pnFourierrep}) for an infinite chain of atoms. The current model of an infinite chain must be treated as an approximation to the more realistic finite chain of $N$ atoms.
In the latter case the wavevector is a finely spaced set of $N$ values. The integral over modes in the nonlinear process then becomes a discrete sum, where it is acceptable to separate a single term in the sum from the other terms. Thus, for the purposes of this iterative calculation it is necessary to treat the integration over modes as a sum over discrete values, whereas for numerical evaluation of final results the integral can be used without any appreciable error for a very long but finite chain. This is similar to situations in quantum optics where a discretisation of modes renders some calculations easier, for example the treatment of thermal radiation~\cite{Loudon}. The RHS terms of $\phi$ in  \eqref{eq:linear_phi_terms} (including those from $Z[\phi]$) in the mode $(k,\omega)$ are set aside, and terms in other modes are iterated using \eqref{eq:linear_phi_terms}.
After $n$ iterations we have a series of terms containing $\phi(k,\omega)$ and powers of $\gamma$ up to $\gamma^n$, with leading order $\gamma^0$ from the linear coupling; terms of order $\gamma^{n+1}$ and higher still contain $\phi$ in different modes. The latter terms are dropped for an approximation to order $\gamma^n$. (Formally, if the iteration process is repeated indefinitely an expression with just $\phi$ in the mode $(k,\omega)$ will result.) When the iteration process is terminated and terms of order $\gamma^{n+1}$ and higher are dropped,  the resulting equation can be written
\begin{align}
\left[
G_\phi
(k,\omega)
\right] &
^{-1}
\phi(k,\omega)
=i\omega \frac{d_0}{\sqrt{a}}
\sqrt{2\pi}\alpha(k)
h(k,\omega)
\nonumber\\
&+
\omega^2
\frac{d_0^2}{a} \left(2\pi|\alpha(k)|^2\right)  
G_p^\prime(k,\omega)
\phi(k,\omega)
,
\label{eq:new_wave_equation}
\\
 & G_p^\prime(k,\omega)=
G_p(k,\omega)
+
O(\gamma)
,
\end{align}
\noindent where $h(k,\omega)$ is now a collection of homogenous solution terms and $G_p$ is now the leading order in a perturbation series giving the new function $G_p^\prime$. We identify $G_p^\prime$ as an effective Green function describing the dressed dipoles $p$ in the nonlinear medium.
We can rewrite this equation in a form similar to \eqref{eq:wave_equation}:
\begin{equation}
\left[
G_\phi^\prime
(k,\omega)
\right]
^{-1}
\phi(k,\omega)
=i\omega \frac{d_0}{\sqrt{a}}
\sqrt{2\pi}\alpha(k)
h(k,\omega),
\end{equation}
\noindent where $G_\phi^\prime$ is the modified Green functions of $\phi$, which can be used to find an effective linear permittivity $\varepsilon_{\textrm{eff}}$:
\begin{equation}
G_\phi^\prime
(k,\omega)
=
\frac{c^2}{
(ck)^2-\omega^2\varepsilon_{\textrm{eff}}(k,\omega)
},
\end{equation}
\begin{equation}
\varepsilon_{\textrm{eff}}(k,\omega)
=1+\frac{d_0^2c^2}{a}\left(2\pi|\alpha(k)|^2\right)
G_p^\prime(k,\omega).
\label{eq:effective_linear_permittivity}
\end{equation}
\section{Diagrams}   \label{sec:diagrams}

The iteration procedure of the previous section rapidly becomes notationally cumbersome.
To simplify this process and the calculation of the effective permittivity, we express the iterative procedure using diagrams. While Feynman diagrams were developed for quantum field theory (QFT) (see~\cite{peskin}, for example), there is nothing inherently quantum about representing a perturbative solution to coupled field equations graphically. Feynman rules for diagrams can also be found when solving classical field equations perturbatively \cite{helling}.

\begin{figure}[htb]
\centering
{\includegraphics[width=60mm]{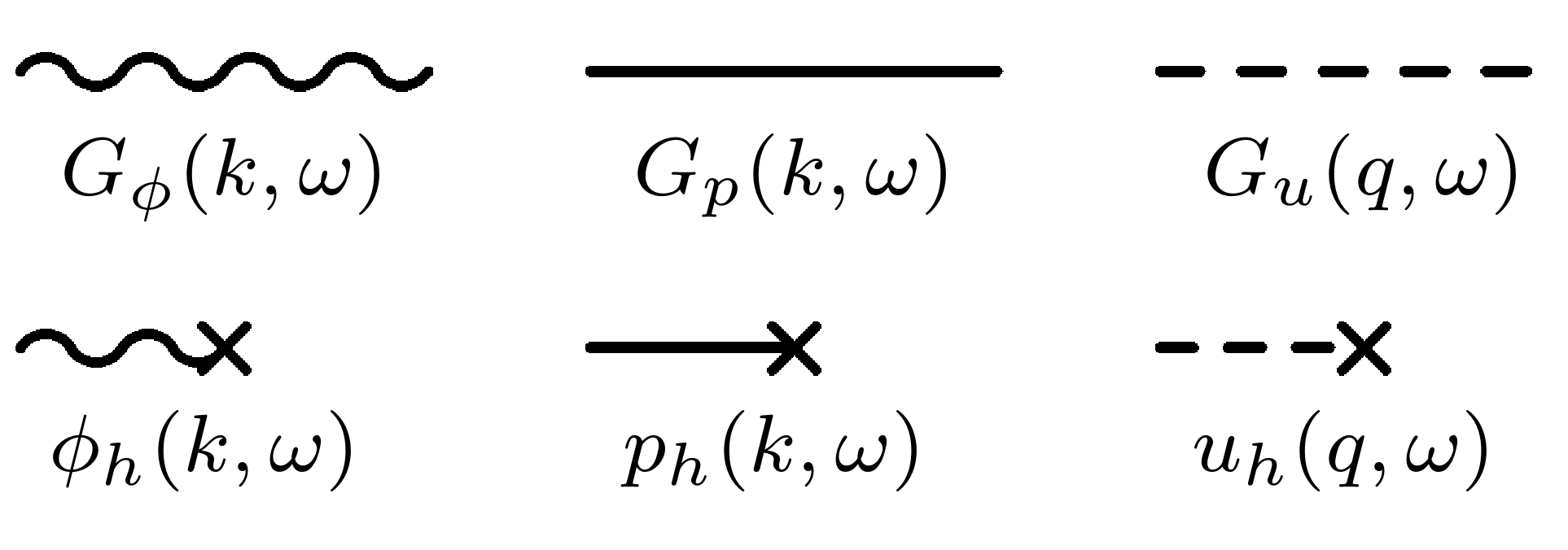}}
\caption{
Diagram representation of Green functions and homogenous solutions in the iteration process.
}
\label{fig:greenfunctions}
\end{figure}

The diagrams are to be read left to right.
After each step in the iteration process, each field is represented as a line: $\phi$, $p$ and $u$ are represented as wavy, straight and dashed lines respectively.
Upon iteration using an equation of motion, a field is replaced by the homogenous solution plus the Green function multiplied by a term involving another field.
For example, in \eqref{eq:nonlinear_u_equation} $u(q,\omega)$ is equal to $u_h(q,\omega)$ plus $G_u(q,\omega)$ multiplied by a $pp$ term that can be iterated further.
As a result all intermediate lines, as shown in Fig.~\ref{fig:greenfunctions}, give a factor of the corresponding Green function, while homogenous solutions that cannot be iterated further are represented as terminated lines.

\begin{figure}[htb]
\centering
{\includegraphics[width=70mm]{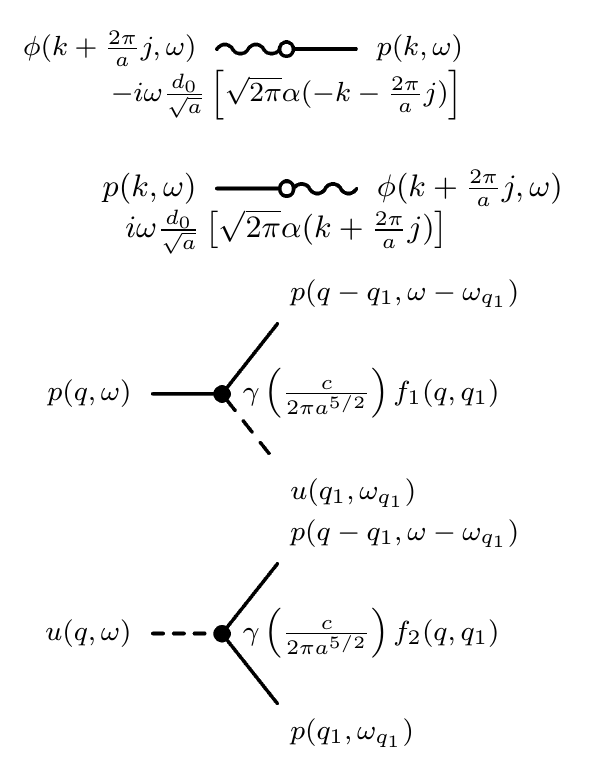}}
\caption{
The allowed vertices and the corresponding coupling factors for the nonlinear model.
}
\label{fig:vertices}
\end{figure}

The lines in a diagram may be connected with a limited number of allowed vertices, determined by the type of coupling in the Lagrangian.
For example, the linear coupling in \eqref{eq:nonlinear_p_equation}  gives a two-line vertex, while the nonlinear coupling gives a three-line vertex.  Each vertex in a diagram has an associated prefactor from the equations of motion.  The vertices and prefactors for the current model are given in Fig.~\ref{fig:vertices}. At each vertex the total frequency of outgoing fields is equal to that of the ingoing field. Due to the periodicity of the material the wavevectors of $p$ and $u$ lie in the first Brillouin zone whereas the wavevector of $\phi$ has no such restriction.  The total wavevector of the outgoing fields at each vertex is equal to that of the ingoing field up to multiples of the reciprocal lattice vector. At a nonlinear vertex, all possible values of the undetermined  $q_1$ and $\omega_{q_1}$ must be integrated over. The main difference to QFT is that we cannot use Wick contraction to close loops and remove additional pairs of $u$, $p$ or $\phi$ terms. As a result, only tree diagrams are permitted.

In summary:
\begin{itemize}
\item	Each intermediate line gives a factor of the corresponding Green function.
\item	Each terminated line gives a factor of the corresponding homogenous solution.
\item	Each vertex gives a factor of the corresponding coupling function from the equation of motion.
\item Frequency $\omega$ is conserved at each vertex. The total wavevector is conserved at each vertex up to multiples of the reciprocal lattice vector, with the restriction that the wavevectors of $p$ and $u$ lie in the first Brillouin zone.  
\item An integral is performed over each undetermined frequency and wavevector variable. 
\item	Only tree diagrams are permitted.
\end{itemize}

This diagrammatic representation gives an intuitive way of finding the modified Green functions in a coupled system, by performing a summation over all diagrams that start and end with the same field. This is very similar to the calculation of the self-energy in QFT. This method also simplifies the identification of terms that can either be grouped together or are part of an infinite series.

As a simple example, we derive $G_p^d$, the Green function of $p$ dressed with $\phi$, using only the linearly coupled model from Sec.~\ref{sec:linear} by performing a summation over all diagrams that start and end with $p(k,\omega)$. In the absence of coupling, we are left with only the first term of the bare Green function. Including the additional diagrams with intermediate $\phi$ steps gives an infinite series that can be expressed via the Dyson equation~\cite{peskin}:
\begin{figure}[htb]
\centering
{\includegraphics[width=70mm]{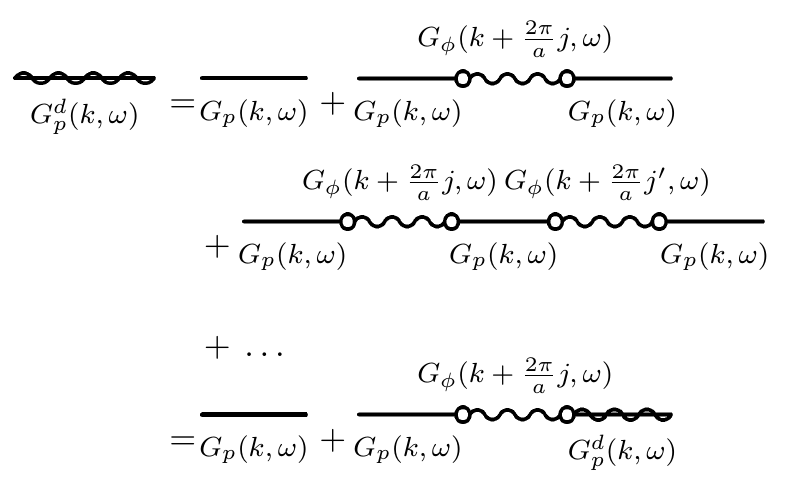}}
\label{fig:dressed_green_function}
\end{figure}
\newpage
Using the rules described, we can evaluate this sum, with each line corresponding to a Green function and each vertex giving a factor of the coupling function:
\begin{align}
G_p^{d}(k,\omega)=&
\phantom{+}
G_p(k,\omega)
\nonumber\\
&+
G_p(k,\omega)
\left[
\sum_{j=-\infty}^{\infty}\omega^2\frac{d_0^2}{a}(2\pi|\alpha(k+\tfrac{2\pi}{a}j)|^2)
\right.
\nonumber\\&
\left.
\phantom{\sum_j^{\infty}}
\times
G_\phi(k+\tfrac{2\pi}{a}j,\omega)
\right]
G_p^{d}(k,\omega).
\end{align}
Dividing by $G_pG_p^d$ gives:
\begin{align}
\left[G_p^{d}(k,\omega)\right]^{-1}=&
\left[G_p(k,\omega)\right]^{-1}
\nonumber\\&-
\sum_{j=-\infty}^{\infty}
\bigg[
\omega^2\frac{d_0^2}{a}(2\pi|\alpha(k+\tfrac{2\pi}{a}j)|^2)
\nonumber\\&
\phantom{-
\sum_{j=-\infty}^{\infty}
\bigg[}
\times
G_\phi(k+\tfrac{2\pi}{a}j,\omega)
\bigg].
\end{align}
Reciprocal lattice vector scattering has been included explicitly in this expression, as although the periodicity of the system restricts the initial wavevector of $p$ to the first Brillouin zone, the intermediate $\phi$ steps are not bound to this condition.
Exactly the same result can be found by substituting \eqref{eq:linear_phi_equation} into \eqref{eq:linear_p_equation}. Using the approximation $\alpha(x)=\delta(x)$ and $\alpha(k)=1/\sqrt{2\pi}$ the sum can be evaluated to give:
\begin{equation}
G_p^{d}(k,\omega)
=
\frac{1}
{\omega_0^2(k)-\omega^2-
\tfrac{d_0^2c}
{2}
\omega
\frac{
\sin(\tfrac{\omega a}{c})
}{
\cos(\tfrac{\omega a}{c})-\cos(ka)
}
},
\label{eq:gdp}
\end{equation}
\noindent where the pole prescription $(\omega+i0^+)$ has again been omitted for notational simplicity.

\begin{figure}[htb]
\centering
{\includegraphics[width=80mm]{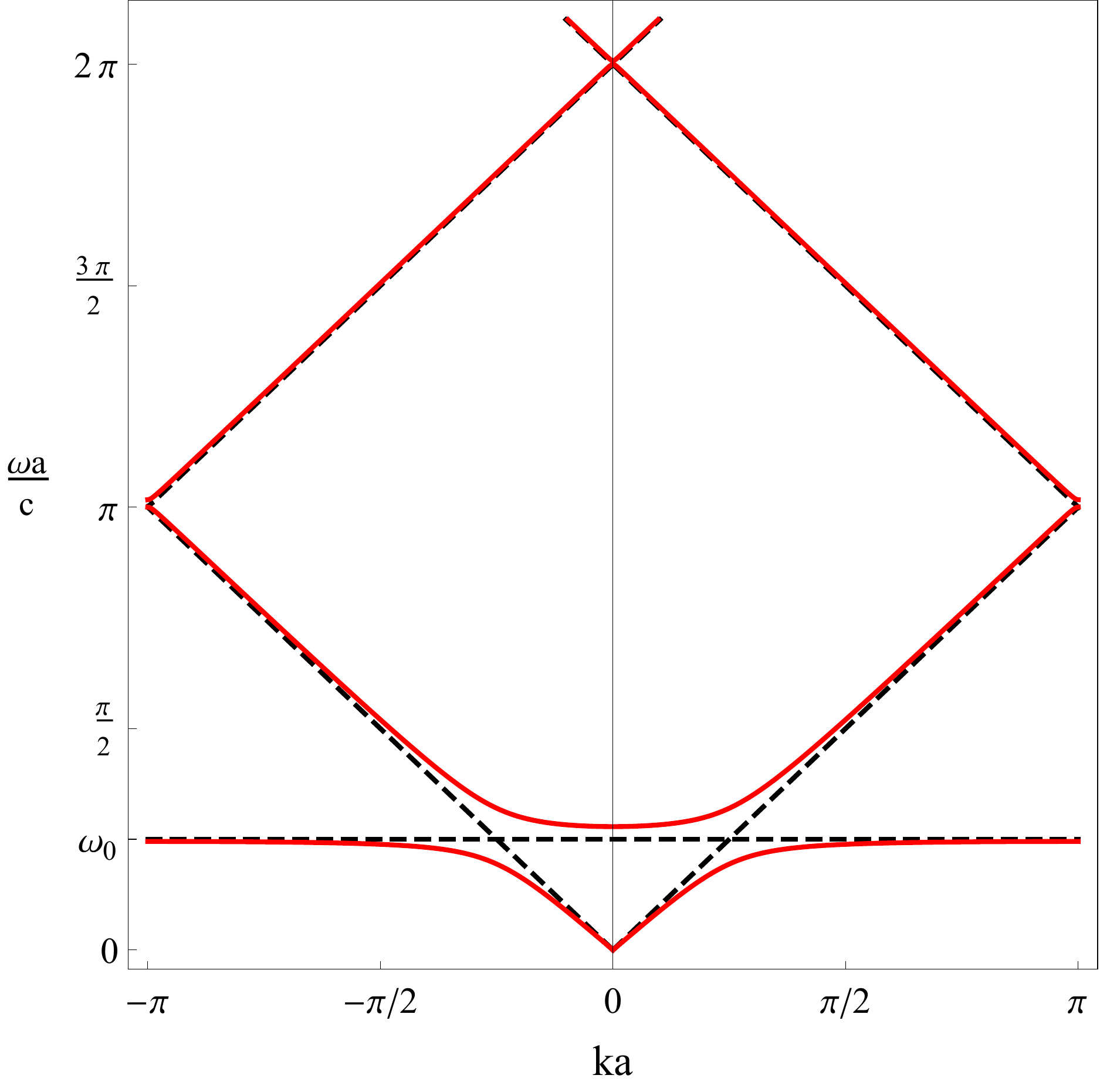}}
\caption{
The dispersion relation for the dressed dipole modes $p$ linearly coupled to a scalar field $\phi$ for a finite $d_0$ (red) compared to the dispersion relations of the uncoupled $p$ and $\phi$ folded back into the first Brillouin zone (dashed black). 
}
\label{fig:dressed_p_dispersion_relation}
\end{figure}

The new dispersion relation for the dressed $p$ is shown in Fig.~\ref{fig:dressed_p_dispersion_relation}. 
The reciprocal lattice vector scattering has the effect of folding the dispersion relation of $\phi$ back into the first Brillouin zone, giving additional branches as $j\to\infty$.
The $\delta$-function approximation of $\alpha(x)$ is accurate for small $\omega$ but may not hold at very large frequencies $(\omega a/c\gg\pi)$ where the dispersion relation is repeatedly folded back into the first Brillouin zone and $j$ becomes large.

\section{Effective Permittivity}   \label{sec:effectiveperm}

\begin{figure}[!htb]\centering
{\includegraphics[width=80mm]{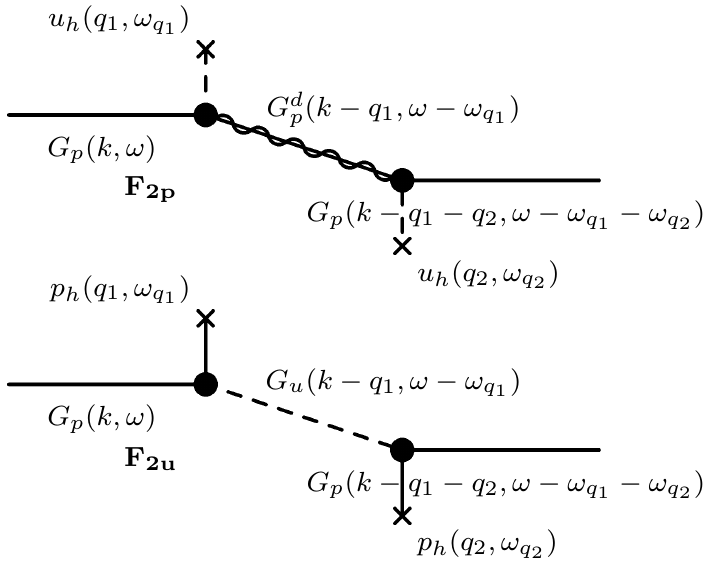}}
\caption{
All diagrams starting and ending with $G_p$ and containing two nonlinear vertices. Note that the dressed Green function $G^d_p$ in \eqref{eq:gdp} is used due to the linear coupling term. Each diagram will give a contribution to $F_2(k,\omega)$; we will denote these contributions by $F_{2p}$ (top) and $F_{2u}$ (bottom), based on the  intermediate step.
}
\label{fig:F_2}
\end{figure}

We now calculate the effective Green function $G_p^\prime(k,\omega)$ in \eqref{eq:new_wave_equation}, which is the Green function of $p$ dressed by the nonlinear interaction; it reduces to the bare Green function $G_p(k,\omega)$ when $\gamma=0$ so linear $p\phi$ vertices only occur between nonlinear vertices. The iteration procedure also  ensures that in $G_p^\prime(k,\omega)$  only intermediate $p$ modes that differ from the ingoing $p$ mode $(k,\omega)$ can couple to $\phi$; intermediate $p$ lines in the ingoing mode $(k,\omega)$ do not connect to $\phi$ lines. The diagrams for $G_p^\prime(k,\omega)$ are thus those that start and end with the bare Green function $G_p(k,\omega)$, have increasing number of nonlinear vertices, and increasing numbers of intermediate $p$ lines in the ingoing mode $(k,\omega)$. 
Similar to the previous section, this can be written as an infinite series:
\begin{figure}[htb]
\centering
{\includegraphics[width=75mm]{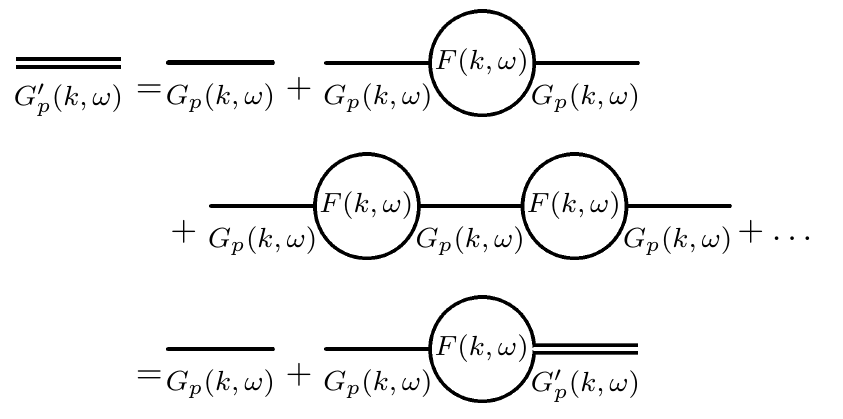}}
\label{fig:Gp_prime}
\end{figure}
\begin{align}
G_p^\prime(k,\omega)
=
G_p(k,\omega)
\bigg\{&
1+
F(k,\omega)G_p(k,\omega)
\nonumber\\
&+
\left[
F(k,\omega)G_p(k,\omega)
\right]^2
+
\dots
\bigg\}
\nonumber\\
=
G_p(k,\omega)&
\left[
1+
F(k,\omega)
G_p^\prime(k,\omega)
\right]
\end{align}
Here, the modified Green function $G_p^\prime$ is represented by a straight double line and $F(k,\omega)$ represents a sum of all diagrams that start and end with $p$ in the mode $(k,\omega)$, where the two outer vertices are nonlinear vertices and the outer $p$ lines are removed. Dividing by $G_pG_p^\prime$ gives:
\begin{equation}
\left[G_p^\prime(k,\omega)\right]^{-1}=
\left[G_p(k,\omega)\right]^{-1}
-
F(k,\omega).
\label{eq:f_sum}
\end{equation}
The complex effective linear permittivity in \eqref{eq:effective_linear_permittivity} is now given by
\begin{equation}
\varepsilon_{\textrm{eff}}(k,\omega)=
1+
\frac{
\left(
d_0^2c^2/a
\right)
\left(2\pi|\alpha(k)|^2\right)
}{
\omega_0(k)^2
-
\omega^2
-
\mathbb{R}\textrm{e}F(k,\omega)
-
i\mathbb{I}\textrm{m}F(k,\omega)
}.
\label{eq:effective_permittivity}
\end{equation}
\noindent Instead of an imaginary Dirac delta function as in \eqref{eq:permittivity}, we have in (\ref{eq:effective_permittivity}) a resonant peak $id_0^2c^2\left(2\pi|\alpha(k)|^2\right)/a\mathbb{I}\textrm{m}F(k,\omega)$ when $\omega_0^2(k)-\omega^2-\mathbb{R}eF(k,\omega)=0$. The complex function $F(k,\omega)$ can be expanded in terms of the number of nonlinear vertices in each diagram: 
\begin{equation}
F(k,\omega)=\gamma^2F_2(k,\omega)+\gamma^3F_3(k,\omega)+\gamma^4F_4(k,\omega)+\dots,
\end{equation}
\noindent where the term $F_n$  contains diagrams where $p$ returns to the initial mode after $n$ scattering processes involving nonlinear vertices and $n-1$ intermediate steps.
Figure \ref{fig:F_2} contains all diagrams that start and end with $G_p$ and contain two nonlinear vertices and corresponding powers of $\gamma$. The dressed Green function $G_p^d$ from \eqref{eq:gdp} has been used to sum over all possible diagrams where the intermediate $p$ step couples to $\phi$ and back any number of times due to the linear coupling term.
The diagram rules can be used to find the corresponding function for each diagram, which will contain an integral over  the possible final modes $(k-q_1-q_2,\omega-\omega_{q_1}-\omega_{q_2})$.
The $F_2$ term will be calculated from the diagrams  in Fig.~\ref{fig:F_2} by isolating the diagrams that return to the initial mode $(k,\omega)$.

We now consider the first diagram in  Fig.~\ref{fig:F_2}.
Using the Feynman rules, the corresponding function is
\begin{widetext}
\begin{align}
G_p(k,\omega)
\bigg\{&
\int^\infty_{-\infty} d\left(\tfrac{\omega_{q_1}a}{c}\right)\,
\int_{-\pi}^\pi d\left(q_1a\right)
\left[
\gamma \left (\frac {c}{2\pi a^{5/2}}\right ) f_1(k,q_1)
\right]
u_h\left(q_1,\omega_{q_1}\right)
 G_p^d(k-q_1,\omega-\omega_{q_1})
\nonumber\\
&\times
\int^\infty_{-\infty} d\left(\tfrac{\omega_{q_2}a}{c}\right)\,
\int_{-\pi}^\pi d\left(q_2a\right)
\left[
\gamma \left (\frac {c}{2\pi a^{5/2}}\right ) f_1(k-q_1,q_2)
\right]
u_h\left(q_2,\omega_{q_2}\right)
G_p(k-q_1-q_2,\omega-\omega_{q_1}-\omega_{q_2})
\bigg\}
,
\label{eq:arbitraryF_2p}
\end{align}
\noindent where we have integrated over all possible final modes.
This expression contains the homogenous solution $u_h$, which must satisfy \eqref{eq:homogenous_u_equation}.
This can be expressed as a delta function in frequency:
\begin{equation}
\frac{u_h\left(q,\omega_{q}\right)c}{2\pi a^{5/2}}=
A(qa)\delta\left(
\frac{\omega_{q} a}{c}-\frac{\Omega(q)a}{c}
\right)
+
A^*(-qa)\delta\left(
\frac{\omega_{q} a}{c}+\frac{\Omega(q)a}{c}
\right),
\end{equation}
\noindent where we have included the preceding factor from the nonlinear vertex to simplify calculations and we have expressed the delta functions in dimensionless variables.
The product $u_h\left(q_1,\omega_{q_1}\right)u_h\left(q_2,\omega_{q_2}\right)$ in (\ref{eq:arbitraryF_2p}) gives four terms, however we only consider those where $\omega_{q_2}=-\omega_{q_1}$, as only these terms allow a return to the initial mode $(k,\omega)$ and thus contribute to $F_2$:
\begin{align}
\left[
\frac{u_h(q_1,\omega_{q_1})c}{2\pi a^{5/2}}
\right]
\left[
\frac{u_h(q_2,\omega_{q_2})c}{2\pi a^{5/2}}
\right]
=\phantom{+}&
A(q_1a)A^*(-q_2a)
\delta\left(
\frac{\omega_{q_1} a}{c}-\frac{\Omega(q_1)a}{c}
\right)
\delta\left(
\frac{\omega_{q_2} a}{c}+\frac{\Omega(q_2)a}{c}
\right)
\nonumber\\
+&
A^*(-q_1a)A(q_2a)
\delta\left(
\frac{\omega_{q_1} a}{c}+\frac{\Omega(q_1)a}{c}
\right)
\delta\left(
\frac{\omega_{q_2} a}{c}-\frac{\Omega(q_2)a}{c}
\right)
+\dots
\label{eq:homogenous_expansion}
\end{align}
Substituting the first two terms of \eqref{eq:homogenous_expansion} into \eqref{eq:arbitraryF_2p} and evaluating the delta functions gives
\begin{align}
G_p(k,\omega)
\bigg\{&
\gamma^2
\int_{-\pi}^\pi d\left(q_1a\right)
f_1(k,q_1)
A\left(q_1a\right)
 G_p^d(k-q_1,\omega-\Omega(q_1))
\nonumber\\
&\times
\int_{-\pi}^\pi d\left(q_2a\right)
f_1(k-q_1,q_2)
A^*\left(-q_2a\right)
G_p(k-q_1-q_2,\omega-\Omega(q_1)+\Omega(q_2))
\bigg\}
\label{eq:arbitraryF_2p_ver2}
\end{align}
\noindent plus another term with $A(qa)\to A^*(-qa)$ and $\Omega(q)\to-\Omega(q)$.
At this point we  separate \eqref{eq:arbitraryF_2p_ver2} into terms that contribute to $F_2$  by returning to the initial $(k,\omega)$ mode  and those that do not.
By considering the integrals as a sum corresponding to a long but finite chain of atoms, we pick out the  $q_2=-q_1$ contribution in the second integral. The expression \eqref{eq:arbitraryF_2p_ver2} then reduces to
$
G_p(k,\omega)
\left[
\gamma^2
F_{2p}\left(k,\omega\right)
\right]
G_p(k,\omega),
$
where $F_{2p}(k,\omega)$ is the desired contribution to $F_{2}(k,\omega)$ and is given by
\begin{align}
F_{2p}(k,\omega)
=\phantom{+}&
\int_{-\pi}^\pi d\left(q_1a\right)
\left|f_1(k,q_1)\right|^2
\left|A\left(q_1a\right)\right|^2
 G_p^d(k-q_1,\omega-\Omega(q_1))
\nonumber\\
+&
\int_{-\pi}^\pi d\left(q_1a\right)
\left|f_1(k,q_1)\right|^2
\left|A\left(-q_1a\right)\right|^2
 G_p^d(k-q_1,\omega+\Omega(q_1))
.
\label{eq:arbitraryF_2p_ver3}
\end{align}
The integrals in (\ref{eq:arbitraryF_2p_ver3}) may contain poles of the dressed Green function $G^d_p$, which coincide with the dispersion relation in Fig.~\ref{fig:dressed_p_dispersion_relation}.
To evaluate \eqref{eq:arbitraryF_2p_ver3} we must insert the $(\omega+i0^+)$ pole prescription in the $G^d_p$  expression \eqref{eq:gdp}, which has the effect of shifting the pole into the lower-half complex $\omega$-plane.
We perform a change in the integration variables from wavevector $q$ to frequency $\Omega(q)$, where it is easier to evaluate the poles:
\begin{equation}
\int_{-\pi}^\pi d\left(qa\right)=
\int_\Omega 
\rho\left(\tfrac{\Omega a}{c}\right)
d\left(\tfrac{\Omega a}{c}\right),
\qquad
\frac{1}{\rho\left(\tfrac{\Omega a}{c}\right)}
=\left|\frac{d\left(\frac{\Omega a}{c}\right)}{d(qa)}\right|,  \label{rhoeq}
\end{equation}
\noindent where $\int_\Omega$ denotes an integral over the finite ranges $-\Omega_{\textrm{max}}a/c$ to $-\Omega_{\textrm{min}}a/c$ and $\Omega_{\textrm{min}}a/c$ to $\Omega_{\textrm{max}}a/c$  using the expression from \eqref{eq:u_resonant_frequency}. The expression (\ref{eq:arbitraryF_2p_ver3}) for $F_{2p}$ now takes the form
\begin{align}
F_{2p}(k,\omega)
=\phantom{+}&
\int_\Omega d\left(\tfrac{\Omega_1 a}{c}\right)
\left|f_1(k,Q(\Omega_1))\right|^2
\left|A\left(\tfrac{\Omega_1 a}{c}\right)\right|^2
\rho\left(\tfrac{\Omega_1 a}{c}\right)
 G_p^d(k-Q(\Omega_1),\omega-\Omega_1)
\nonumber\\
+&
\int_\Omega d\left(\tfrac{\Omega_1 a}{c}\right)
\left|f_1(k,-Q(\Omega_1))\right|^2
\left|A\left(\tfrac{\Omega_1 a}{c}\right)\right|^2
\rho\left(\tfrac{\Omega_1 a}{c}\right)
 G_p^d(k+Q(\Omega_1),\omega-\Omega_1)
,
\label{eq:arbitraryF_2p_ver4}
\end{align}
\end{widetext}
\noindent where $Q(\Omega)$ is the inverse function of $\Omega(q)$, with the properties $Q(\Omega)=Q(-\Omega)$ and $Q(\Omega)>0$. 

The functional form of the lattice amplitude $|A(\Omega_1 a/c)|$ in (\ref{eq:arbitraryF_2p_ver4}) must be specified. It is natural to take the homogeneous solution for the lattice as a thermal state. The average amplitude of a classical harmonic oscillator in thermal equilibrium is inversely proportional to its frequency so we take
\begin{equation}
\left|A\left(\tfrac{\Omega a}{c}\right)\right|^2
=
A_0^2
\left(
\frac{\Omega_{\textrm{min}}}{\Omega}
\right)^2,
\label{eq:A0}
\end{equation}
\noindent where $A_0$ is dimensionless.

As in the linear case \eqref{eq:G_p_expansion}, the pole prescription of the retarded Green function $G^d_p$ in (\ref{eq:arbitraryF_2p_ver4}) can be used to split the integrals into real principal value integrals plus imaginary terms associated with the poles. The latter terms correspond to the dressed dipole mode  lying on the dispersion relation of Fig.~\ref{fig:dressed_p_dispersion_relation} and can be found analytically in terms of residues; they give the imaginary part of (\ref{eq:arbitraryF_2p_ver4}) as
\begin{widetext}
\begin{align}
\mathbb{I}\textrm{m}F_{2p}(k,\omega)
=
&
i\pi \sum_n \textrm{Res}
\left[
\left|
f_1(k,Q(\Omega_{1}))
\right|^2
\left|
A\left(\tfrac{\Omega_{1}a}{c}\right)
\right|^2
\rho\left(-\tfrac{\Omega_{1}a}{c}\right)
G_p^{d}(k-Q(\Omega_{1}),\omega-\Omega_{1})
,
\Omega_{\textrm{pole}, n}
\right]
\nonumber\\
+&
i\pi \sum_m \textrm{Res}
\left[
\left|
f_1(k,-Q(\Omega_{1}))
\right|^2
\left|
A\left(-\tfrac{\Omega_{1}a}{c}\right)
\right|^2
\rho\left(\tfrac{\Omega_{1}a}{c}\right)
G_p^{d}(k+Q(\Omega_{1}),\omega-\Omega_{1})
,
\Omega_{\textrm{pole}, m}
\right]
,
\label{eq:ImF2p}
\end{align}
\end{widetext}
\noindent where $\Omega_{\textrm{pole}, n}$ is the $n^{th}$ pole in the $\Omega_{1}$ integration (the range of which is described after (\ref{rhoeq})) and $\textrm{Res}[f(z),z_0]$ denotes the residue of $f(z)$ at $z_0$. The principal value integrals that give the real part of (\ref{eq:arbitraryF_2p_ver4}) must be calculated numerically. An additional check on this numerical calculation can be made by comparing with the result obtained by using the Kramers-Kronig relations on the imaginary part (\ref{eq:ImF2p}) of $F_{2p}$.

In addition to $F_{2p}$, there is another contribution to $F_2$, associated with the second diagram in Fig.~\ref{fig:F_2}, and which we label $F_{2u}$. This contributes very differently, however, for the following reason. The linear coupling between $p$ and $\phi$ means the poles of the intermediate Green function $G^d_p$ in $F_{2p}$, which coincide with the dispersion relation in  Fig.~\ref{fig:dressed_p_dispersion_relation}, occur for almost all frequency arguments in $G^d_p$. 
This gives a nonzero imaginary part (\ref{eq:ImF2p}) of $F_{2p}(k,\omega)$ for nearly every $\omega$ (recall that the residues in this equation are for the finite frequency range described after (\ref{rhoeq})).
In contrast, the poles in the intermediate Green function $G_u$ in $F_{2u}$ (see second diagram in Fig.~\ref{fig:F_2}) occur when $\omega^2=\Omega^2(q)$. The corresponding dispersion relation runs over a smaller, finite range of frequencies determined by \eqref{eq:u_resonant_frequency}.
This smaller range means that the first few leading order $F$ terms with intermediate $u$ steps will not have a pole in the integral over intermediate modes for a large range of initial frequencies $\omega$.
For example, in the region of interest near the dipole resonant frequency $\omega=\omega_0$, $F_{2u}$ has no imaginary component for the choice of model parameters made in the next section.
For this reason we only consider diagrams with intermediate dressed $p$ steps in the numerical calculations that follow.

The $F_{2p}$ expression (\ref{eq:arbitraryF_2p_ver4}) contains complex conjugate pairs of the homogenous solutions and the vertex coupling function.
The integration over intermediate states gives a constructively adding quantity as a result of this. Higher order terms in  $F$ do not necessarily have such complex conjugate pairs, instead containing a mixture of homogenous solutions and vertex functions at different frequencies and wavevectors. Upon integration, these can interfere destructively.
In calculating the higher order terms, we only retain diagrams that give complex conjugate pairs, which give constructive interference and the dominant contributions to $F$.
The next terms that satisfy this condition belong to the $F_4$ group and are calculated using the same process as $F_{2p}$. The diagram and corresponding expression for the leading order term $F_{4p}$ is given in the appendix.
Both  $F_{2p}$ and $F_{4p}$ will be used in the following numerical calculations, as they are the leading contributions to $F$.

\section{Numerical Calculations}   \label{sec:numerical}

For numerical calculations, we consider a lattice spacing $a\approx3$\AA\ and use the approximation $\alpha(k)=1/\sqrt{2\pi}$ for small initial $k$ values within the first Brillouin zone. The resonant frequency $\omega_0$ is taken to be in the visible region at $477.44\,\mathrm{THz}$ ($1.975\,\mathrm{eV}$), which corresponds to the dimensionless quantity $\omega_0a/c=0.003$.
The coupling term $\tau_{1} (a/c)^2=1\times10^{-9}$ in \eqref{eq:resonant_frequency} is chosen so that $\omega_0(k)$ is approximately constant in $k$.
The lattice dispersion relation $\Omega(q)$ covers a typical frequency range for a solid of approximately $0.1\to10\,\mathrm{THz}$ \cite{wei1994phonon}. 
The dimensionless terms $A_0$ and $\gamma (a/c)^2$ are chosen so that the $F$ sum is convergent and perturbation theory is valid.
For the purpose of our calculations $A_0$ and  $\gamma (a/c)^2$ are $0.0003$ and $1\times10^{-7}$ respectively.
Our general formulas are not specific to these values and are valid provided that the convergent $F$ sum condition is satisfied.

The most important constant is the linear coupling term $d_0$, which determines the size of the gap near $\omega=\omega_0$ in the dressed $p$ dispersion relation, shown in Fig.~\ref{fig:dressed_p_dispersion_relation} (larger $d_0$ gives a larger gap).
If $d_0$ is too large, the finite frequency integral in $F_{2p}$ over intermediate modes for an initial frequency $\omega\approx\omega_0$  will not include any poles of the intermediate Green function $G^d_p$ or the corresponding imaginary residue terms.
If $d_0$ is too small, the integral includes poles from both the upper and lower branches of the dispersion relation in Fig.~\ref{fig:dressed_p_dispersion_relation}, with the possibility of the imaginary residues cancelling each other.
In both of these cases, one of the higher order terms such as $ F_{4p}$ will dominate the imaginary part of the $F$ perturbation series.
For the purposes of our calculations, we consider the intermediate case with $d_0\sqrt{a}=0.003$ , where the integral over intermediate modes for an initial frequency $\omega\approx\omega_0$ only includes the lower branch of the dispersion relation and $\mathbb{I}\textrm{m}F_{2p}$ is nonzero.

\begin{figure}[htb]{}
\centering
{\includegraphics[width=80mm]{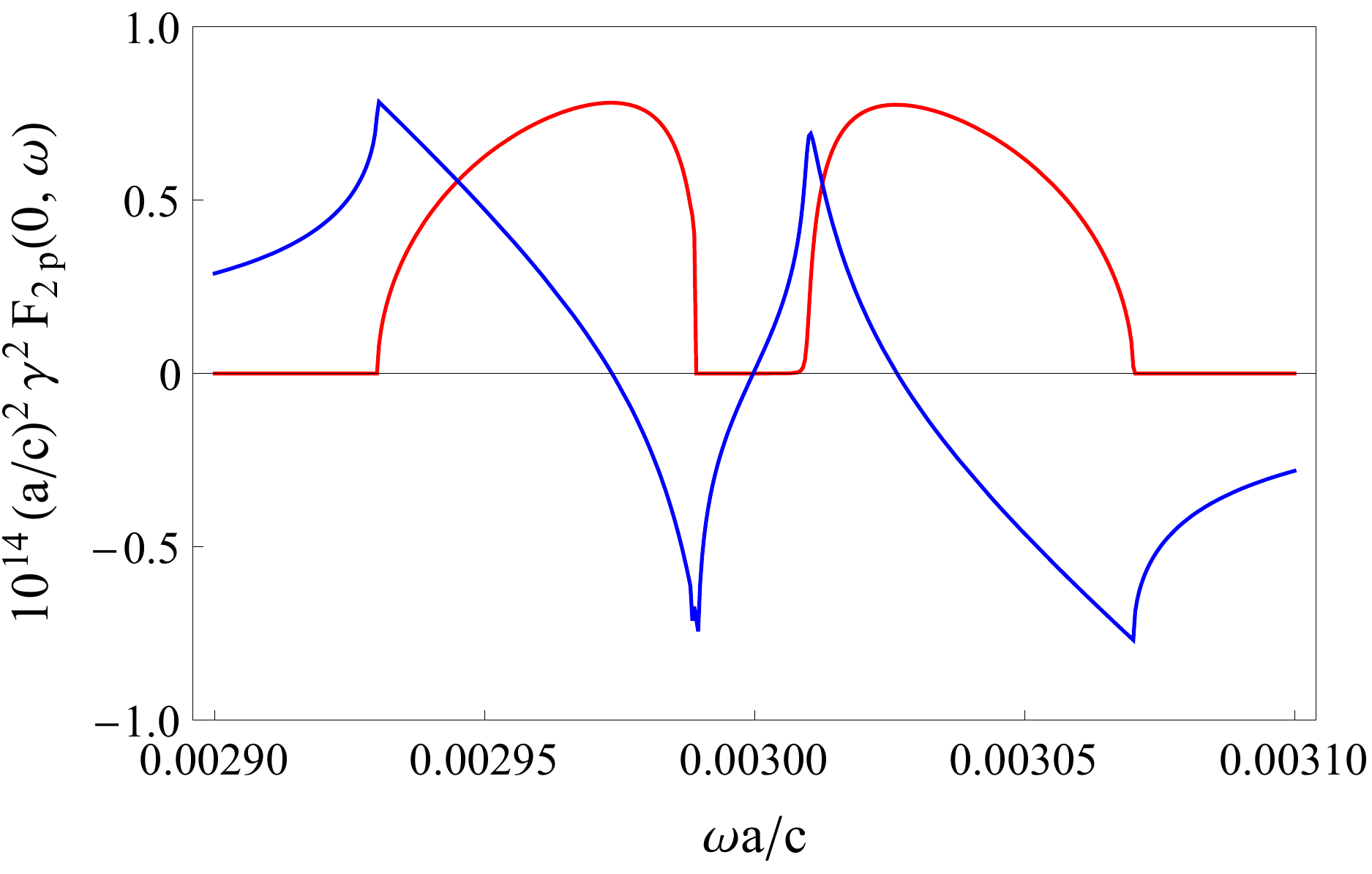}}
\caption{$\mathbb{R}\textrm{e}F_{2p}$ (blue) and $\mathbb{I}\textrm{m}F_{2p}$ (red) rescaled to dimensionless variables in the frequency region near $\omega_0(k)$ for $k=0$.
The real and imaginary parts are related by the Kramers-Kronig relations.
}
\label{fig:F2_Com}
\end{figure}

We first consider $k=0$, where $F_{2p}$ dominates $F$ for the chosen values of $A_0$ and $\gamma$; in this $k=0$ case the $F_{4p}$ contribution is not significant. Figure \ref{fig:F2_Com} shows the real and imaginary parts of $F_{2p}$ near the dipole resonant frequency $\omega_0$. The peaks in the imaginary part occur when the pole in the intermediate Green function $G^d_p$ lies on the ``flat'' part of the lower branch of the dispersion relation in Fig.~\ref{fig:dressed_p_dispersion_relation}. 
The shape of the peaks is determined by $f_1(k,\omega)$ and $A(\Omega a/c)$ from \eqref{eq:f1} and \eqref{eq:A0}. The value of  $\mathbb{I}\textrm{m}F_{2p}(0,\omega)$ between the two peaks near $\omega_0$ is small, but nonzero. As expected, the real and imaginary parts obey the Kramers-Kronig relations.

\begin{figure}[htb]{}
\centering
{\includegraphics[width=80mm]{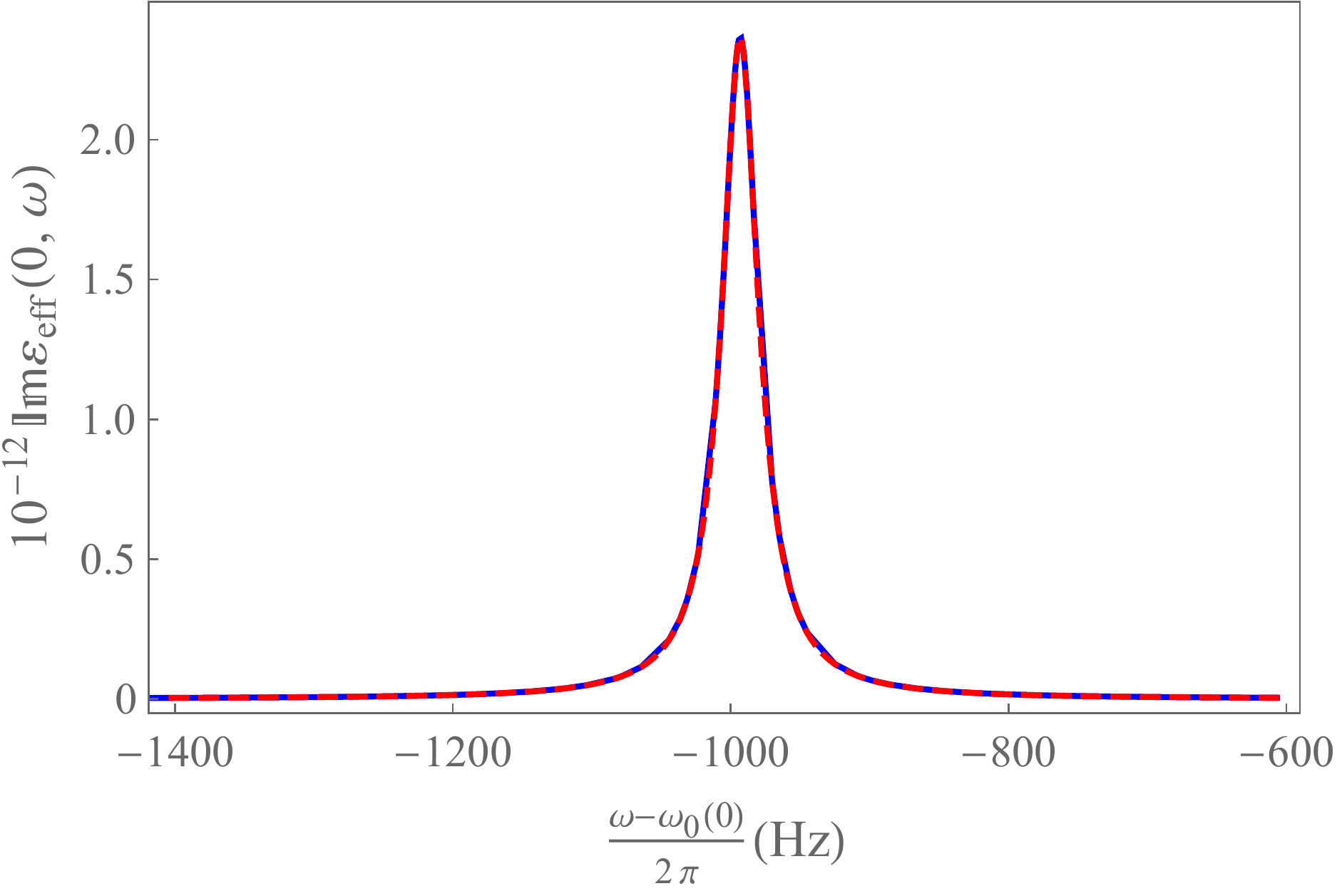}}
\caption{
The imaginary permittivity (\ref{eq:imaginary_permittivity}) (blue solid) and the DHO model fit (\ref{epD}) (red dashed) when $k=0$. The resonant peak has moved slightly from the position of the Dirac delta function of the linear model by $\sim1\,\mathrm{kHz}$.
The calculated permittivity is extremely well fitted by the DHO model.
}
\label{fig:imaginary_permittivity}
\end{figure}

From \eqref{eq:effective_permittivity} the imaginary part of the permittivity is
\begin{equation}
\mathbb{I}\textrm{m}\varepsilon_{\textrm{eff}}(k,\omega)=
\frac{
\left(
d_0^2c^2/a
\right)
\left[
\mathbb{I}\textrm{m}F(k,\omega)
\right]
}{
\left[
\omega_0(k)^2
-
\omega^2
-
\mathbb{R}\textrm{e}F(k,\omega)
\right]^2
+
\left[
\mathbb{I}\textrm{m}F(k,\omega)
\right]^2},
\label{eq:imaginary_permittivity}
\end{equation}
\noindent which has a resonant peak when $\omega_0^2(k)-\omega^2=\mathbb{R}eF(k,\omega)$.
The small value of $\gamma$ and the shape of $\mathbb{R}eF(k,\omega)$ mean the resonant frequency will only be shifted slightly from $\omega_0(k)$.
The size of the peak is determined by $\mathbb{I}\textrm{m}F(k,\omega)$.
Figure \ref{fig:imaginary_permittivity} shows the resonant peak in $\mathbb{I}\textrm{m}\varepsilon_{\textrm{eff}}$ for $k=0$.
The small values of $\gamma$ and $A_0$ give the peak a small linewidth of $\sim60\,\mathrm{Hz}$ and a large maximum value of $2.4\times10^{12}$.
The peak is only slightly shifted from the resonant value by $\sim1\,\mathrm{kHz}$, corresponding to a $2\times10^{-12}$ fractional shift. The peaks in $\mathbb{I}\textrm{m}F_{2p}$ in Fig.~\ref{fig:F2_Com} also give features either side of the central resonant peak in Fig.~\ref{fig:imaginary_permittivity}. These are  smaller than the central peak by several orders of magnitude as the resonant condition $\omega_0^2(k)-\omega^2=\mathbb{R}eF(k,\omega)$ is not satisfied. The extremely sharp peak in Fig.~\ref{fig:imaginary_permittivity} is due to our use of perturbation theory with a very small value of the nonlinear coupling parameter $\gamma$. A larger $\gamma$ would require more terms in the perturbation series to be evaluated, which involve more complicated intermediate scattering processes. More realistic results for the permittivity would require consideration of a very large number of intermediate processes, and permittivity values comparable to those measured in real dielectrics may be beyond the scope of perturbation theory in our model. However, we have shown that Hopfield's proposal~\cite{Hopfield} is correct: nonlinear interactions between the dipoles and lattice vibrations act as a pseudo-reservoir giving an effective permittivity with finite imaginary part. The resulting functional form of the permittivity is also in line with Hopfield's statement~\cite{Hopfield} that such nonlinear material interactions should produce a permittivity agreeing with the standard damped harmonic oscillator (DHO) model
\begin{equation}
\varepsilon_D(\omega)
=1+\frac{A_D}{\omega_D^2-\omega^2-i\gamma_D\omega}.  \label{epD}
\end{equation}
\noindent Figure~\ref{fig:imaginary_permittivity} shows a fit of the imaginary part of the permittivity to the imaginary part of (\ref{epD}), where $A_D$, $\omega_D$ and $\gamma_D$ take the values $A_D=1.43\times10^{30}\,\mathrm{s}^{-2}$, $\omega_D=\omega_0-6240\,\,\mathrm{s}^{-1}$ and $\gamma_D=202\,\mathrm{s}^{-1}$.
It is readily apparent that our model gives an extremely good fit to the DHO formula (\ref{epD})  commonly used to describe real dielectrics. This close agreement is due to the fact that the imaginary term in the denominator of (\ref{epD}) ($\gamma_D\omega$) and nonlinear model (\ref{eq:effective_permittivity}) ($\mathbb{I}\textrm{m}F(k,\omega)$) do not change significantly over the width of the peak. This imaginary term can thus be approximated as a constant, reducing both expressions to the Lorentzian function.
The real part of the permittivity agrees with the DHO model to a similar degree, obeying the Kramers-Kronig relations.
This behaviour is not specific to the nonlinear coupling used here and including more complex nonlinear coupling terms in the Lagrangian also gives a resonant peak that is an excellent fit to a DHO model, albeit with different parameters.

\begin{figure}[htb]{}
\centering
{\includegraphics[width=86mm]{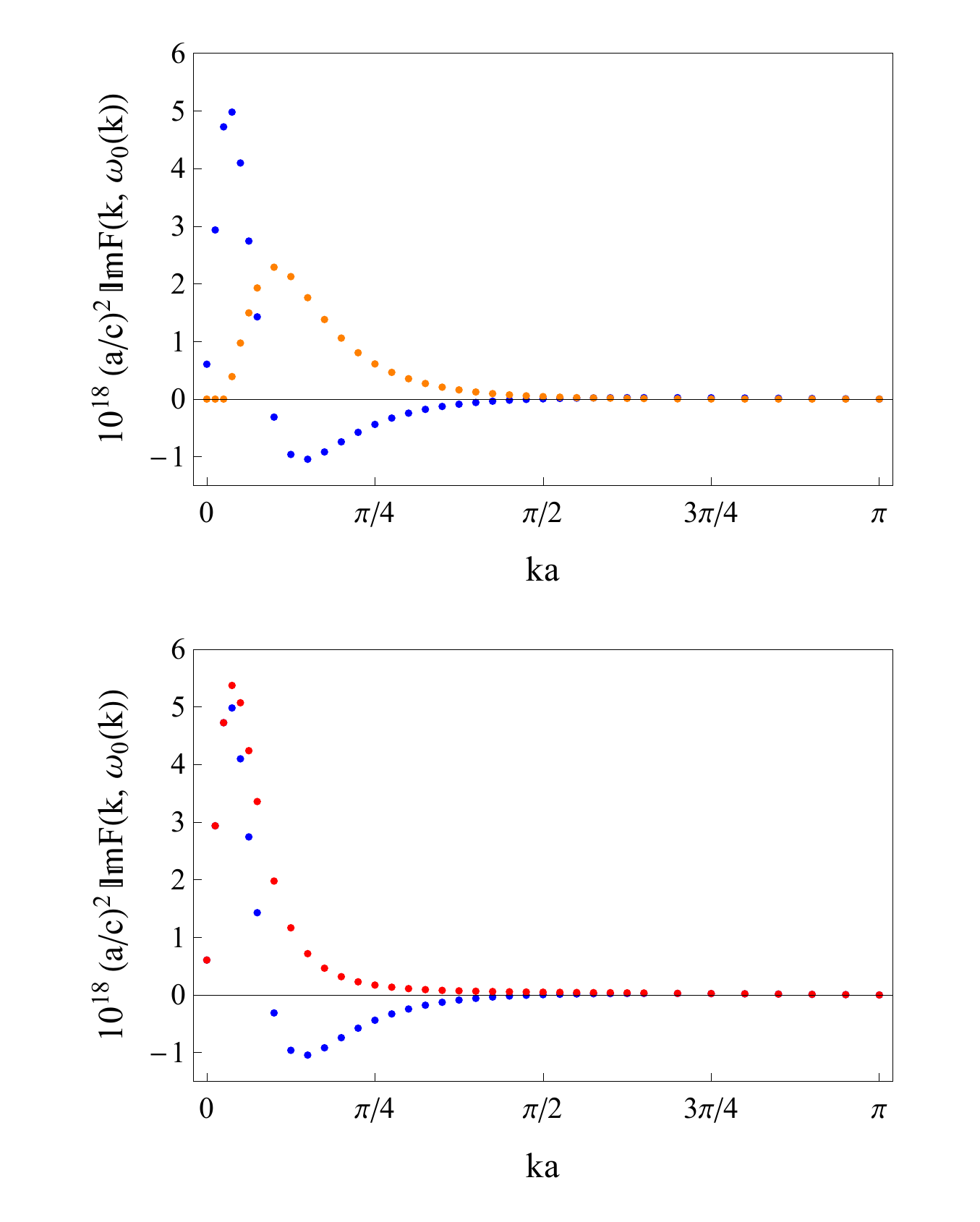}}
\caption{
Top:
$\gamma^2\mathbb{I}\textrm{m}F_{2p}(k,\omega_0(k))$ (blue) and $\gamma^4\mathbb{I}\textrm{m}F_{4p}(k,\omega_0(k))$ (orange) as a function of $k$.
Bottom:
$\mathbb{I}\textrm{m}F(k,\omega_0(k))$ as a function of $k$, comparing the first ($F_{2p}$) term (blue) to the sum of the $F_{2p}$ and $F_{4p}$ terms (red).
$F_{2p}$ is the dominant term for small $k$ within the light cone.
At larger wavevectors, the residues in $F_{2p}$ begin to cancel and higher order terms become dominant in the $F$ perturbation series.
}
\label{fig:imaginary_F_side_by_side}
\end{figure}

\section{Spatial dispersion}   \label{sec:spatial}

We now consider a nonzero wavevector in $F(k,\omega)$ to investigate spatial dispersion in the medium, i.e.\ the wavevector dependence of the permittivity (\ref{eq:effective_permittivity}).
The first plot in Fig.~\ref{fig:imaginary_F_side_by_side} shows the behaviour of the imaginary pats of the two leading terms $F_{2p}$ and $F_{4p}$ of $F$ at the dipole resonant frequency $\omega_0(k)$, as a function of $k$. The second plot in Fig.~\ref{fig:imaginary_F_side_by_side} shows $\mathbb{I}\textrm{m}F(k,\omega_0(k))$ with just the $F_{2p}$ contribution included and then also with the $F_{4p}$ contribution added.
For small $k$, $F_{2p}(k,\omega_0(k))$  gives the dominant contribution to $\mathbb{I}\textrm{m}F(k,\omega_0(k))$. In this case the only processes in the dominant $F_{2p}$ term that give a residue in \eqref{eq:ImF2p} are those from an integral containing poles from the lower branch of the dressed $p$ dispersion relation.
There are two such poles, with opposite signs of wavevector.
As $k$ increases, the residue term from the pole with the opposite sign of wavevector to $k$ increases, while the residue from the pole with the same sign of wavevector as $k$ decreases and changes sign. The overall value of $F_{2p}$ decreases and can become negative depending on the model parameters and coupling, as is seen in Fig.~\ref{fig:imaginary_F_side_by_side}.
The higher order $F_{4p}$ term of $F$ offsets the negative $F_{2p}$ term in Fig.~\ref{fig:imaginary_F_side_by_side}, with $F_{4p}$ then dominating the expression for $\mathbb{I}\textrm{m}F$.
The full $k$-dependence as $ka\to\pi$ would require calculating many terms and is not pursued further here.

As $k$ increases, the peak in the imaginary part of the permittivity (\ref{eq:effective_permittivity}), shown in Fig.~\ref{fig:imaginary_permittivity} for $k=0$, decreases and broadens. This effect can be modelled by the DHO formula (\ref{epD}) by replacing the parameters with a power series in $k$, for example
\begin{equation}
\omega_D(k)=\omega_{D0}+\omega_{D2}k^2+\omega_{D4}k^4+\dots, \label{oDexp}
\end{equation}
where only even powers are present due to the symmetry of the system. For the numerical values used in this section, a $k^2$ expansion in (\ref{oDexp}) provides a good fit to the calculated permittivity for $k$ up to $ka=0.1$, with further terms in (\ref{oDexp}) required for higher $k$. A $k^2$ expansion in (\ref{oDexp}), giving a $k^2$ term in the denominator of (\ref{epD}), was proposed by Hopfield and Thomas~\cite{hopfield1963theoretical} based on different considerations. Information on the spatial dispersion of materials is limited in comparison to temporal dispersion, and the former is usually treated as of minor importance~\cite{LLcm}. Nevertheless an understanding of spatial dispersion is essential for accurate predictions in the nano-optics of small particles~\cite{raz11,wie12,pen12} and also for the prediction of Casimir and thermal forces on an isolated object~\cite{horsley}. Results from our model may help clarify how spatial dispersion in dielectrics operates over a significant range of wavevectors.

\section{Conclusions}
We have developed a simple classical model of a dielectric that features nonlinear interactions between polarizable ``atoms" and lattice vibrations. Our motivation was to verify the main claims of Hopfield~\cite{Hopfield} regarding this model. Results such as those presented here give a better quantitative understanding of the mechanism of light absorption in dielectrics, and also provide information on spatial dispersion. The lattice vibrations act as a pseudo-reservoir into which  electromagnetic energy is dissipated and the resulting permittivity is closely approximated by the standard textbook formula (\ref{epD}). As well as dissipation of incident radiation into the medium, our model will also describe re-emission of radiation out of the medium once the lattice is excited. The latter process is not captured by the effective permittivity and is contained in nonlinear $\phi$ terms in our perturbation procedure that were not analysed here.

We note that in our classical calculations the lattice must already be excited (we chose a thermally excited state as the coupling-independent part of the lattice solution) in order to perform its reservoir role. A quantization of the model would presumably also give broadband absorption of light at zero temperature due to the zero-point energy of the lattice. 

The reservoir role of the nonlinearly coupled lattice is captured at a phenomenological level by linear coupling to a continuum reservoir of harmonic oscillators at all frequencies~\cite{Huttner}. The continuum reservoir, linearly coupled to the electromagnetic field, is in turn sufficient to give a Lagrangian formulation of the macroscopic Maxwell equations for arbitrary materials obeying Kramer-Kronig relations\cite{bha06,khe06,sut07,amo08,amo09,khe10,philbin2010canonical}. 

Spatial dispersion emerges naturally from our model. The wavevector dependence of the effective permittivity shows agreement with simpler considerations~\cite{hopfield1963theoretical} for small $k$, but our model allows the calculation of higher-order contributions that are necessary for a full characterisation of the nonlocal response.

The model explored here may also find application in microwave metamaterials. Dipoles with a sharp resonance and very low internal loss that are arranged in a lattice that can vibrate could serve as a macroscopic system that is well described by our model. 

\acknowledgements
We acknowledge financial support from EPSRC under Program Grant EP/I034548/1.

\section*{APPENDIX: HIGHER ORDER $F$ TERMS}
\label{sec:appendix}

\begin{figure}[htb]
\centering
{\includegraphics[width=86mm]{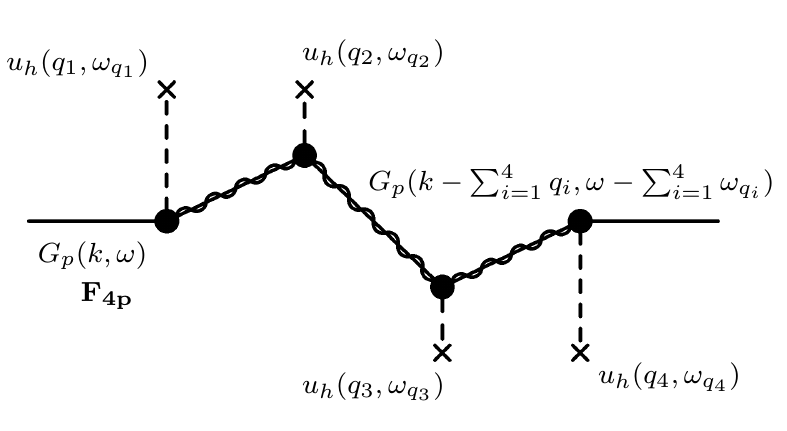}}
\caption{
Diagram starting and ending with $G_p$ and containing four nonlinear vertices, which gives a contribution to $F_4$, which we denote $F_{4p}$ based on the intermediate steps. The additional restrictions on higher order $F$ terms  to  give complex conjugate pairs of homogenous solutions and vertex coupling functions reduce the middle portion to the $F_{2p}$ diagram.
}
\label{fig:F_4}
\end{figure}

The next term in the $F$ sum after $F_2$ that satisfies the additional conditions in Sec.~\ref{sec:effectiveperm} (requiring complex conjugates of vertex functions and homogenous solutions)  belongs to the $F_4$ group. While this contains many diagrams, we consider the term $F_{4p}$ calculated using the diagram in Fig.~\ref{fig:F_4} and named after the intermediate $p$ steps.

As before, the expression for the diagram in Fig.~\ref{fig:F_4} is found using the Feynman rules and the expression for $F_{4p}$ is found by only considering terms that return to the initial mode $(k,\omega)$.
Due to the extra conditions we have imposed on the higher order $F$ terms, we choose $u(q_3,\omega_{q_3})=u(-q_2,-\omega_{q_2})$ and $u(q_4,\omega_{q_4})=u(-q_1,-\omega_{q_1})$ to give complex conjugate pairs of the  homogenous solutions and  vertex functions. In this case the middle step of the diagram is now the same as $F_{2p}$. The final expression for $F_{4p}$ can be reduced to
\begin{widetext}
\begin{align}
F_{4p}(k,\omega)
=
\int_{\Omega} d\left(\tfrac{\omega_{q_1}a}{c}\right)
\bigg\{
&
\left|
f_{1}(k, Q(\omega_{q_1}))
\right|^2
\left|
A\left(\tfrac{\omega_{q_1}a}{c}\right)
\right|^2
\rho\left(-\tfrac{\omega_{q_1}a}{c}\right)
\left[
G_p^{d}(k- Q(\omega_{q_1}),\omega-\omega_{q_1})
\right]^2
\nonumber\\
&
\times
\left[
F_{2p}(k-Q(\omega_{q_1}),\omega-\omega_{q_1})
\right]
\bigg\}
\nonumber\\
+
\int_{\Omega} d\left(\tfrac{\omega_{q_1}a}{c}\right)
\bigg\{
&
\left|
f_{1}(k,-Q(\omega_{q_1}))
\right|^2
\left|
A\left(-\tfrac{\omega_{q_1}a}{c}\right)
\right|^2
\rho\left(\tfrac{\omega_{q_1}a}{c}\right)
\left[
G_p^{d}(k+Q(\omega_{q_1}),\omega-\omega_{q_1})
\right]^2
\nonumber\\
&
\times
\left[
F_{2p}(k+Q(\omega_{q_1}),\omega-\omega_{q_1})
\right]
\bigg\}
.
\label{eq:F_4pp}
\end{align}
\end{widetext}
The singularities in (\ref{eq:F_4pp}) are dealt with in the same manner as those in $F_{2p}$, by splitting the integral into a principal value integral and a residue term.



\begin{thebibliography}{99}
\bibitem{jac}
J.\ D.\ Jackson, {\it Classical Electrodynamics,} 3rd ed.\ (Wiley, New York, 1999).

\bibitem{Hopfield}
J.\ Hopfield, Phys.\ Rev.\ {\bf 112}, 1555 (1958).

\bibitem{Tait}
W.\ C.\ Tait and R.\ I.\ Weiner, Phys.\ Rev.\ {\bf 166}, 769 (1968); {\bf 178}, 1404 (1969).

\bibitem{Hiz}
V.\ V.\ Hizhnyakov, Phys.\ Stat.\ Sol. {\bf 34}, 421 (1969).

\bibitem{Mav}
C.\ Mavroyannis, J.\ Math.\ Phys. {\bf 11}, 491 (1970).

\bibitem{Egl}
W.\  Egler and H.\ Haken, Z.\ Physik B {\bf 28}, 51 (1977).

\bibitem{Cam}
H.\ N.\ Cam, N.\ V.\ Hieu and N.\ A.\ Viet, Phys.\ Stat.\ Sol. (B) {\bf 126}, 247 (1984).

\bibitem{Huttner}
B.\ Huttner and S.\ M.\ Barnett, Phys.\ Rev.\ A {\bf 46}, 4306 (1992).

\bibitem{Suttorp}
L.\ G.\ Suttorp and M.\ Wubs, Phys.\ Rev.\ A {\bf 70}, 013816 (2004).

\bibitem{bha06}
N.\ A.\ R.\ Bhat and J.\ E.\ Sipe, Phys.\ Rev.\ A {\bf 73} 063808 (2006).

\bibitem{khe06}
F.\ Kheirandish and M.\ Amooshahi, Phys.\ Rev.\ A {\bf 74} 042102 (2006).

\bibitem{sut07}
L.\ G.\ Suttorp, J.\ Phys.\ A {\bf 40} 3697 (2007).

\bibitem{amo08}
M.\ Amooshahi and F.\ Kheirandish, J.\ Phys.\ A {\bf 41} 275402 (2008).

\bibitem{amo09}
M.\ Amooshahi, J.\ Math.\ Phys.\ {\bf 50} 062301 (2009).

\bibitem{khe10}
F.\ Kheirandish, M.\ Soltani and J.\ Sarabadani, Phys.\ Rev.\ A {\bf 81} 052110 (2010).

\bibitem{philbin2010canonical}
T.\ G.\ Philbin, New J.\ Phys.\ {\bf 12}, 123008 (2010); {\bf 13}, 063026 (2011).

\bibitem{philbin2012quantum}
T.\ G.\ Philbin, New J.\ Phys.\ {\bf 14}, 083043 (2012).

\bibitem{zha09}
Q. Zhao, J. Zhou, F. Zhang and D. Lippens, Mater. Today {\bf 12}, 60 (2009).

\bibitem{sou11}
C. M. Soukoulis and M. Wegener, Nat. Photonics {\bf 5}, 523 (2011).

\bibitem{sch07}
J. A. Schuller, R. Zia, T. Taubner and M. L. Brongersma, Phys. Rev. Lett. {\bf 99}, 107401 (2007).

\bibitem{wan14}
C. Wang, Z. Y. Jia, K. Zhang, Y. Zhou, R. H. Fan, X. Xiong and R. W. Peng, J. Appl. Phys. {\bf 115}, 244312 (2014).

\bibitem{jia16}
Z. Y. Jia, J. N. Li, H. W. Wu, C. Wang, T. Y. Chen, R. W. Peng and M. Wang, J. Appl. Phys. {\bf 119}, 074302 (2016).

\bibitem{mag59}
V.\ B.\ Magalinskii, Sov.\ Phys.\ JETP {\bf 9} 1381 (1959).

\bibitem{fey63}
R.\ P.\ Feynman and F.\ L.\ Vernon, Ann.\ Phys.\ {\bf 24} 118 (1963).

\bibitem{cal83}
A.\ O.\ Caldeira and A.\ J.\ Leggett, Physica A {\bf 121} 587 (1983).

\bibitem{Tatarskii}
V.\ I.\ Tatarski\v{\i}, Sov.\ Phys.-Usp.\ {\bf 30}, 134 (1987).

\bibitem{Loudon}
R.\ Loudon, {\it The Quantum Theory of Light,} 3rd ed.\ (Oxford university Press, Oxford, 2000).

\bibitem{peskin}
M.\ E.\ Peskin and D.\ V.\ Schroeder, {\it An Introduction to Quantum Field Theory} (Addison-Wesley, 1995)

\bibitem{helling}
R.\ C.\ Helling,
\url{homepages.physik.uni-muenchen.de/~helling/classical_fields.pdf}

\bibitem{wei1994phonon}
S.\ Wei and M.\ Chou, Phys.\ Rev.\ B {\bf 50}, 2221 (1994).

\bibitem{hopfield1963theoretical}
J.\ Hopfield and D.\ Thomas, Phys.\ Rev.\ {\bf 132}, 563 (1963).

\bibitem{LLcm}
L.\ D.\ Landau, E.\ M.\  Lifshitz and L.\ P.\ Pitaevskii, {\it Electrodynamics of Continuous Media} 2nd ed.\ (Butterworth-Heinemann, Oxford, 1984).

\bibitem{raz11}
S.\ Raza, G.\ Toscano, A.\ P.\ Jauho, M.\ Wubs and N.\ Asger Mortensen, Phys. Rev. B\ {\bf 84} 121412(R) (2011).

\bibitem{wie12}
A.\ Wiener, A.\ I.\ Fern\'andez--Domi\'inguez, A.\ P.\ Horsfield, J.\ B.\ Pendry and S.\ A.\ Maier, Nano Lett. {\bf 12} 3308 (2012).

\bibitem{pen12}
A.\ I.\ Fern\'andez--Domi\'inguez, A.\ Wiener, F.\ J.\ Garc\'ia--VIdal, S.\ A.\ Maier S A and J. B.\ Pendry, Phys. Rev. Lett. {\bf 108} 106802 (2012).

\bibitem{horsley}
S.\ A.\ R.\ Horsley and T.\ G.\ Philbin, New J.\ Phys.\ {\bf 16}, 013030 (2014).

\end{thebibliography}
\end{document}